\documentclass[useAMS,usenatbib]{mnras}

\usepackage{newtxtext,newtxmath,xcolor}

\usepackage[T1]{fontenc}
\usepackage{ae,aecompl}
\usepackage{multirow}
\usepackage{caption}
\hypersetup{draft}

\usepackage{graphicx,mathtools}	
\usepackage{amsmath}	
\usepackage{amssymb}	

\title[Solar-wind models for pulsar timing]{On the usefulness of existing Solar-wind models for pulsar timing corrections}

 \author[C. Tiburzi et al.]{C.~Tiburzi,$^{1,2,3}$\thanks{E-mail: ctiburzi@mpifr-bonn.mpg.de}
 J.~P.~W.~Verbiest,$^{2,1}$
 G.~M.~Shaifullah,$^{3}$
 G.~H.~Janssen,$^{3,4}$
 \newauthor
 J.~M.~Anderson,$^{5}$
  A.~Horneffer,$^{1}$
  J.~K\"unsem\"oller,$^{2}$
  S.~Os\l{}owski,$^{6}$
  \newauthor
  J.~Y.~Donner,$^{1,2}$
  M.~Kramer,$^{1,7}$
  A.~Kumari,$^{8}$
  N.~K.~Porayko,$^{1}$
  P.~Zucca,$^{3}$
  \newauthor
  B.~Ciardi,$^{9}$ 
  R.-J.~Dettmar,$^{10}$
  J.-M.~Grie\ss{}meier,$^{11,12}$
  M.~Hoeft,$^{13}$
  M.~Serylak$^{14,15,11}$
 \\
 \\
 $^{1}$Max-Planck-Institut f\"ur Radioastronomie, Auf dem H\"ugel 69, 53121 Bonn, Germany\\
 $^{2}$Fakult\"at f\"ur Physik, Universit\"at Bielefeld, Postfach 100131, 33501 Bielefeld, Germany\\
 $^{3}$ASTRON, the Netherlands Institute for Radio Astronomy, Oude Hoogeveensedijk 4, 7991 PD, Dwingeloo, the Netherlands\\
 $^{4}$Department of Astrophysics/IMAPP, Radboud University, P. O. Box 9010, 6500 GL Nijmegen, The Netherlands\\
 $^{5}$Deutsches GeoForschungsZentrum GFZ, Telegrafenberg, 14473 Potsdam, Germany\\
 $^{6}$Swinburne University of Technology, PO Box 218, Hawthorn, VIC 3122, Australia\\
  $^{7}$Jodrell Bank Centre for Astrophysics, School of Physics and Astronomy, The University of Manchester, Manchester M13 9PL, UK\\
  $^{8}$Indian Institute of Astrophysics, 2nd Block, Koramangala, Bangalore 560034, India\\
  $^{9}$Max-Planck-Institut f\"ur Astrophysik, Karl-Schwarzschild-Stra\ss{}e 1, D-85748 Garching b. M\"unchen, Germany\\
  $^{10}$Fakult\"at f\"ur Physik und Astronomie, Astronomisches Institut, 44780 Bochum, Germany\\
  $^{11}$LPC2E - Universit\'{e} d'Orl\'{e}ans /  CNRS, 45071 Orl\'{e}ans cedex 2, France\\
  $^{12}$Station de Radioastronomie de Nan\c{c}ay, Observatoire de Paris, PSL Research University, CNRS, Univ. Orl\'{e}ans, OSUC, \\{ }18330 Nan\c{c}ay, France\\
  $^{13}$Th\"uringer Landessternwarte, Sternwarte 5, 07778 Tautenburg, Germany\\
  $^{14}$South African Radio Astronomy Observatory, 2 Fir Street, Black River Park, Observatory, Cape Town, 7925, South Africa\\
 $^{15}$Department of Physics and Astronomy, University of the Western Cape, Cape Town 7535, South Africa
 }

\date{Accepted XXX. Received YYY; in original form ZZZ}

\pubyear{2018}

\begin{document}
\label{firstpage}
\pagerange{\pageref{firstpage}--\pageref{lastpage}}
\maketitle

\begin{abstract}
Dispersive delays due to the Solar wind introduce excess noise in high-precision pulsar timing experiments, and must be removed in order to achieve the accuracy needed to detect, e.g., low-frequency gravitational waves. In current pulsar timing experiments, this delay is usually removed by approximating the electron density distribution in the Solar wind either as spherically symmetric, or with a two-phase model that describes the contributions from both high- and low-speed phases of the Solar wind. However, no dataset has previously been available to test the performance and limitations of these models over extended timescales and with sufficient sensitivity. Here we present the results of such a test with an optimal dataset of observations of pulsar~J0034$-$0534, taken with the German stations of LOFAR. We conclude that the spherical approximation performs systematically better than the two-phase model at almost all angular distances, with a residual root-mean-square (rms) given by the two-phase model being up to 28\% larger than the result obtained with the spherical approximation. Nevertheless, the spherical approximation remains insufficiently accurate in modelling the Solar-wind delay (especially within 20~degrees of angular distance from the Sun), as it leaves timing residuals with rms values that reach the equivalent of 0.3~$\mu$s at 1400~MHz. This is because a spherical model ignores the large daily variations in electron density observed in the Solar wind. In the short term, broadband observations or simultaneous observations at low frequencies are the most promising way forward to correct for Solar-wind induced delay variations.

\end{abstract}




\section{Introduction}\label{sec:intro}

 Pulsars are highly magnetised, rapidly rotating neutron stars \citep{hbp+68} that generate co-rotating beams of broadband radiation, most easily observable at radio wavelengths. When such a beam of radiation passes through the line of sight (LoS) of an observer, it appears as a pulsed signal. The periodic arrival of pulsar radiation at Earth can be precisely timed, in particular for millisecond pulsars \citep[MSPs,][]{bkh82}. This property makes MSPs excellent probes for a range of experiments, e.g., to perform tests of theories of gravity through high-precision timing experiments \citep[for a review see][]{wex14}. Among the most exciting projects in the field are \textit{Pulsar Timing Arrays} \citep[PTAs; see e.g.][]{fb90,vlh16,hd17,tib18}, that aim to detect gravitational waves (GWs) in the nanohertz regime by timing an array of carefully selected MSPs. It is most likely that the first GW signal to be detected by these experiments is an isotropic and stochastic gravitational wave background (GWB, see e.g. \citealt{rsg15}).\\

Along the propagation path, pulsar radiation passes through several ionised media: the ionised interstellar medium (IISM), the Solar wind (SW), and the Earth's ionosphere, and each of these affect the propagation of radio waves by introducing a dispersive delay $\Delta t$. Such delays are well described by the cold-plasma dispersion relation, and are proportional to the electron column density and the inverse square of the observing frequency $f$:

\begin{equation}\label{eq:dispersivedelay}
 \Delta t = \mathcal{D} \frac{\rm DM}{f^2},
\end{equation}

 where $\mathcal{D}\approx4.15 \times 10^3$~MHz$^2$~pc$^{-1}$~cm$^3$~s \citep{lk04}, and DM is the \textit{dispersion measure}, defined as the path integral along the LoS of the free electron density. The DM can vary as the LoS moves in the sky to track a pulsar motion \citep[see e.g.][]{hlk+04}, and DM variations are of particular concern for all studies that aim to achieve high precision in pulsar timing, such as PTAs. \\
 For most pulsars, the major contribution to DM variations comes from the turbulent and inhomogeneous IISM, that can induce fluctuations up to a few $10^{-3}$~pc/cm$^{3}$ on the timescale of years \citep[see e.g.][]{jml17} because of pulsars' non-negligible transverse velocities \citep[see e.g.][]{dcl16,mnf16}. \\
 The SW \citep[for a review see][]{schw06} introduces the next most significant contribution. The mean DM due to the SW is $\sim$10$^{-4}$~pc/cm$^{3}$ at a Solar angle (i.e., the angular distance between the pulsar and the Sun, as seen from Earth) of $\sim$60~degrees and it decreases with angular distance from the Sun. The delay induced by the SW varies on timescales of days, Solar rotations (27 days), years and the Solar activity cycle ($\sim$11 years).\\ 
 The DM contribution of the ionosphere is in the order of $10^{-5}$~pc/cm$^{3}$, and it is negligible with respect to the current sensitivities. In this article, we will thus focus on separating and analysing the DM variations caused by the IISM and the SW.\\
  
Fluctuations in DM are typically modelled as arising from spatial structures in the electron density which pass through the line of sight (LoS) because of the relative motions of the pulsar, the Earth and the intervening plasma. Note that temporal variations in the density structure of the plasma would also induce DM variations, but they are not appreciable because their propagation speed is slower than the LoS velocity. For example, the propagation speed of information in the SW can be approximated with the Alfv\'{e}n speed, that is lower than $10^3$~km$/$s at Solar distances larger than a few Solar radii \citep{wm05,zcb14}. The dominant contribution to the LoS speed is the rotational velocity of Earth (approximately 4$\times10^{-3}$~degrees per second), that can be translated into a linear velocity of $\sim10^5$~km$/$s near the Sun, which exceeds the Alfv\'{e}n speed in the SW by an order of magnitude. Similar considerations are valid in the IISM, and hence temporal variations are not a concern.\\

In both cases, it is assumed that the density variations are turbulent in origin, but in neither case the turbulence is homogeneous on the entire LoS.\\ 
Within the IISM, the mean DM is an integral over hundreds of parsecs, but the density fluctuations are often dominated by smaller regions of higher density around $\sim$10 astronomical units (AU) in size \citep[see e.g.][]{fdjh87,cks15}, while in the SW there are several distinct structures: slow dense streams, fast diffuse streams, \textit{co-rotating interaction regions} (CIRs) where a fast stream overtakes a slow stream, and distinct plasmoids ejected from the Sun and carried out by the SW called \textit{coronal mass ejections} (CMEs). The streams and CIRs are quasi-static, persisting for several Solar rotations. The CMEs are one-time transients and are difficult to predict, but are easily seen in white-light coronagraph polarized brightness images.\\
 
 While advanced analyses on IISM-induced DM variations on pulsar signal have been carried out for years \citep[see e.g.][]{rtd88,kcs13,jml17}, to assess their impact on high-precision pulsar timing, studies on the SW effects have not been regularly revised. One of the most recent SW analyses, presented by \citealt{mca19}, attempts to model the SW contribution in regions far from the Sun using a spherical harmonic decomposition which is constrained to the zeroeth order. In their model, the ISM contribution is modelled as a slowly varying contribution from a turbulent Kolmogorov phase screen at some distance from the Earth. This contribution is removed by means of a low-pass filtering scheme, and the remaining DM variation is attributed to the SW and a joint analysis is carried out for 45 pulsars to detect time and helio-latitude dependent structure in the SW. However, the authors suggest that their data are insensitive to the temporal variations of the SW, and the most constrained model appears to be the static in time model for which they recover a mean electron density at 1 astronomical unit (AU) of $\sim$7.9~cm$^{-3}$. In general, no SW model in pulsar timing has even been tested against sensitive low-frequency data. \\
 
 In this article we use a high-cadence, four-year long dataset of PSR~J0034$-$0534 \citep{bhl+94} observed at $\sim$150~MHz with four German stations of LOFAR \citep[the LOw Frequency ARray,][]{vwg13} to (1) properly test the SW models available in pulsar timing, and (2) assess their usefulness for high-precision pulsar timing experiments such as PTAs. 
 In Section~\ref{sec:modeldescription}, we describe the SW and the available SW models in pulsar timing. In Section~\ref{sec:observations} we introduce the telescopes used and the observing setup, and in Section~\ref{sec:datared} the data reduction and the method applied to disentangle the IISM effects from those due to the SW. In Section~\ref{sec:evaluation} we focus on evaluating the performance of the two models, while in Section~\ref{sec:ptaimpact} we discuss the implications of our finding for high-precision pulsar-timing experiments. In Section~\ref{sec:conclusions} we summarise our conclusions.
 
\section{The Solar wind and pulsar timing models}\label{sec:modeldescription}

\subsection{Solar wind structure}\label{sec:swstructure}

The SW is a flow of magnetised plasma that originates from the Solar corona and continues outwards at constant velocity to a distance of $\sim$100~AU where it terminates. It is much better observed than the IISM, since we have 50 years of direct spacecraft observations and as many years of remote sensing by radio propagation methods. At Solar angles larger than 10~degrees, the SW velocity is roughly constant, and thus the density falls like distance squared because of the mass conservation principle. Although the density near the Sun is quite variable spatially, a rough bi-modality can be detected \citep{col96}: a high-velocity (600 to 800~km/s) and low-density mode ($\sim$3~cm$^{-3}$), called the \textit{fast wind}, versus a low-velocity ($<$400~km/s, see \citealt{tkf10}) and denser mode ($\sim$5 to 10~cm$^{-3}$, see \citealt{man03}) called the \textit{slow wind}. The 11-year Solar activity cycle is an important factor in the SW behaviour. At the minimum of Solar activity, an open magnetic field region (called a \textit{coronal hole}) is located over each of the poles. While streams of fast wind are emitted from the coronal holes, around the Solar equator there is a belt of slow wind, where the neutral magnetic field line (i.e., the locus of points where the Solar magnetic field is zero) is embedded. As the Solar activity increases, the slow-wind belt tilts to follow the magnetic-pole shift, away from the rotational pole. This allows the fast wind to appear in the ecliptic plane (which is tilted at 7.5~degrees with respect to the rotational equator). As the Solar activity increases, the coronal hole fragments and slow wind is seen at all latitudes. Consequently, it is difficult to see a latitude effect integrated over a solar cycle, but it is a dominant feature during the minimum phase of the activity cycle. This behaviour was discovered with radio scintillation events and later confirmed by the Ulysses spacecraft \citep{psb13}.\\

 As the SW propagates outwards, the rotation of the Sun will cause fast-wind streams to move under previously emitted slow-wind ones. Such an interaction causes a compression region that increases with distance from the Sun. While a uniform slow-wind flow at Earth would have a density of about 10~cm$^{-3}$, in a compression region this value can double. These phenomena are the already mentioned CIRs \citep[see, e.g., ][]{gp99, ric18}, that are particularly prominent at the minimum phase of the Solar activity cycle. In addition to the spatial variations of fast streams, slow streams, and CIRs, there are CMEs \citep[for a review see][]{che11}. CMEs are transient events in which a loop of magnetic field projecting outwards from the Sun becomes unstable and ejects a compact plasmoid carrying with it a closed loop of magnetic field. They are relatively common phenomena, that happen about one or two times per day or more depending on the phase of the Solar cycle, and are emitted with an initial diameter of a Solar radius to then expand linearly with distance. The chances of observing one during a three-hour observation are not large, but they cannot be ignored. This is because, whether a CME transits in front of a pulsar, it will introduce a DM excess that is unpredicted by the models, and the affected observation would appear as an outlier.

\subsection{One-phase SW model}\label{sec:t2model}
The most commonly used pulsar timing packages, \textsc{tempo}\footnote{\url{http://tempo.sourceforge.net/}} and \textsc{tempo2}\footnote{\url{http://tempo2.sourceforge.net/}, see also \citet{hem06,ehm06}}, offer a built-in model to mitigate the SW influence. This model assumes a constant SW speed and preserved mass flux, and thus describes the free electron density in the SW as purely spherical and decreasing with the square of the radial distance $R$ from the Sun:

\begin{equation}\label{eq:sphericalmodel}
 n_{\rm e} (R) = {n_0} \langle \frac{R_0}{R} \rangle^2,
\end{equation}

where $n_0$ is the free electron density of the SW at the Earth, $R_0$ is the distance between the Sun and the Earth, and both $R_0$ and $R$ are expressed in Solar radii. \textsc{tempo} and \textsc{tempo2} differ with respect to the $n_0$ value, that is set at 9.961 and 4~cm$^{-3}$ respectively. However, the $n_0$ value assumed by \textsc{tempo} is an extremely high value, even to model the DM effects at the Solar maximum, while the value used in \textsc{tempo2} (hereafter T2) is reasonable for most of the activity cycle, but low for the Solar maximum phase. While both software packages presently allow the user to choose the density at Earth to optimally match their data (which effectively means that the amplitude of this model is free), $n_0$ is a variable parameter in reality, and neither model allows for any spatial variation other than the radial gradient.\\

While the spherical approximations are the most widely used SW models in pulsar timing, it is clear that they are unable (by design) to encompass the high-frequency temporal fluctuations of the SW, and to provide any latitude-dependent correction or information about SW transients. 
\subsection{Two-phase SW model}
\citet[hereafter Y07]{yhc07}, presented a more detailed model that accounts for the slow- and the fast-wind phases of the large-scale structure of the SW. In this model, the electron density of each phase is described by independent scaling laws that depend on the radial distance from the Sun. Specifically, for the slow wind the authors propose:

 \begin{equation}\label{eq:slowwind}
\begin{multlined}
 n_{\rm e, slow} =  2.99 \times 10^{14} R^{-16} + 1.5\times10^{14}R^{-6} + \\  + 4.1\times10^{11} ( R^{-2} + 5.74R^{-2.7})~ m^{-3},
\end{multlined}
 \end{equation}

while for the fast wind:

 \begin{equation}\label{eq:fastwind}
\begin{multlined}
 n_{\rm e, fast} =  1.155 \times 10^{11} R^{-2} + 32.3\times10^{11}R^{-4.39} + \\  + 3254 \times10^{11}  R^{-16.25}~ m^{-3}.
\end{multlined}
 \end{equation}

In the Y07 model, the SW speed and the latitude range around the neutral magnetic field line (where the slow wind originates) are fixed at 400~km/s (for both slow and fast wind) and 20~degrees respectively.  \\

 \begin{figure}
   \includegraphics[width=\columnwidth]{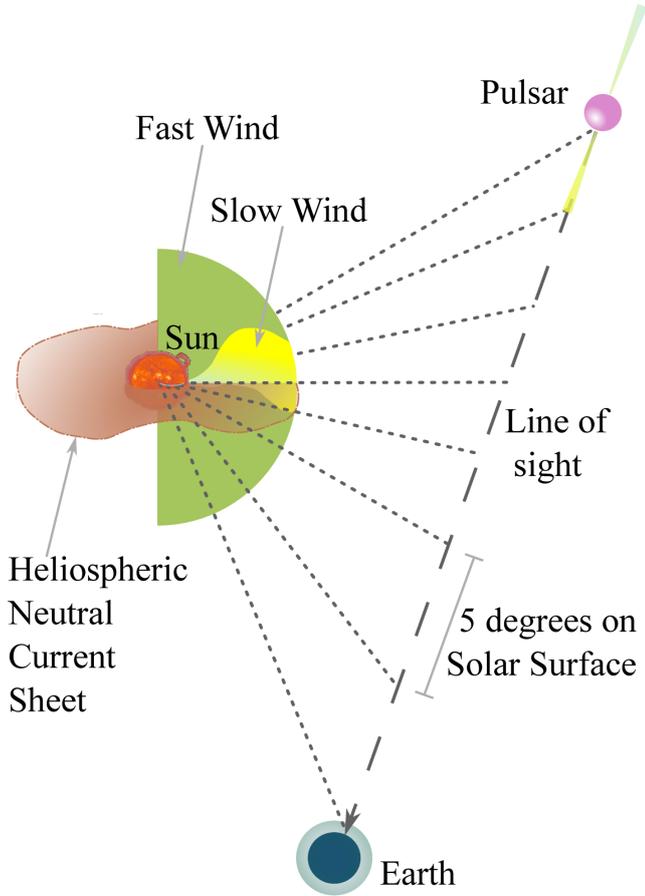}
  \caption{Cartoon that shows the basic functioning of the Y07 model. The slow and fast phases of the SW, and the neutral line are shown in yellow, green, and brown respectively. The dashed line indicates the LoS towards the pulsar observed from Earth, while the dotted lines isolate segments of 5~degrees as subtended from the Sun. Each segment is then back-projected onto the Solar surface.}
   \label{fig:y07model_sketch}
 \end{figure}
 
\begin{figure*}
  \includegraphics[scale=0.55, trim= 1cm 0 4cm 0]{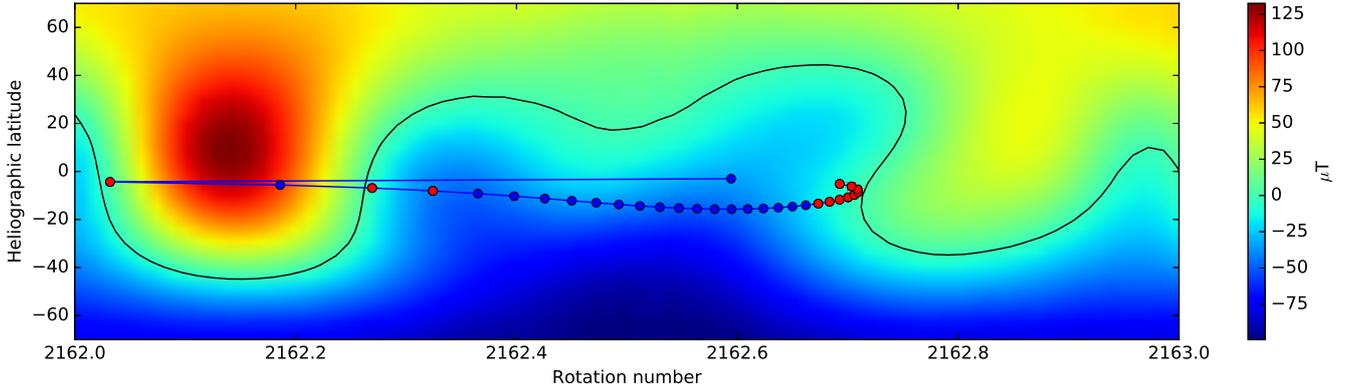}
  \caption{Synoptic chart of the coronal magnetic field (shown as a colour intensity diagram in $\mu$T, and obtained with the CSS extrapolation scheme) at MJD~57132 as a function of Carrington rotation (i.e., the number of Solar rotations, assumed to be ${\sim}27.28$ days long each, since the 9th of November 1853).}
    \label{fig:coronalmap}
 \end{figure*}

To compute the expected dispersive contribution due to the SW for a certain pulsar, Y07 divides the LoS into segments which subtend 5 degrees at the Solar surface (see Figure~\ref{fig:y07model_sketch}). Each segment is back-projected onto magnetic field maps of the Solar corona through the assumed SW speed (see Figure~\ref{fig:coronalmap}). This allows for an inference on the position of each segment with respect to the neutral field line, and hence the SW phase that affects them and the corresponding column density of free electrons (as given by equations~\ref{eq:slowwind},~\ref{eq:fastwind}). The column densities of all the segments are then summed and scaled to the units of DM. The heliographic coordinates of the neutral line can be obtained from the synoptic charts of the coronal magnetic field provided by the Wilcox Solar Observatory (WSO)\footnote{The synoptic charts can be found at: \url{http://wso.stanford.edu/forms/prsyn.html}. To download the chart, select the desired number of Carrington rotation or the start date. Select ``ClassicCSS'' as extrapolation scheme, then ``Final'' as field type, and ``Latitude'' as projection.}. It should be noted, however, that such maps cannot be measured directly. Therefore, these maps are extrapolated from the (directly observed) magnetic field of the photosphere\footnote{The WSO offers three types of extrapolation schemes: CSS (that assumes a non-radial magnetic field, and that the Solar corona is located at 2.5~R$_{\odot}$), R250 (assumes a radial magnetic field, and that the Solar corona is located at 2.5~R$_{\odot}$), and R350 (as R250, but assuming the location of the Solar corona to be at 3.5~R$_{\odot}$). More information can be found at: \url{http://wso.stanford.edu/synsourcel.html} and references therein.}, by assuming the space between the photosphere and the corona to be current-free. In this article, we use synoptic maps of the Solar corona obtained with the CSS extrapolation scheme (as in \citealt{yhc07})\footnote{Note that we obtain the same results by using any of the interpolation schemes.}.\\

\section{Observations}\label{sec:observations}
The international LOFAR telescope is an array of sub-arrays, referred to as \textit{stations}, that operates between 30 and 240~MHz. Each station is made of two different types of antenna, one type for the low part of the band (30 to 80~MHz) and the other for the high part (110 to 240~MHz). 38 stations are located in the Netherlands, and a further 13 sub-arrays are international and cover longer baselines. The international stations can be used independently as stand-alone telescopes -- in particular, the six international stations in Germany (referred to as DE601, DE602, DE603, DE604, DE605, and DE609) are operated by members of the German Long-Wavelength (GLOW) consortium\footnote{\url{https://www.glowconsortium.de/}} and are mainly used to carry out a long-term, high-cadence monitoring campaign of pulsars.\\

The continuing pulsar monitoring campaign observes more than 100 pulsars weekly in the high band (\citealt{mhd18}, \citealt{sto18}, \citealt{pnt19}, \citealt{dvt19}). The signal is digitised at the station and channelised by a polyphase filterbank into 195 kHz-wide subbands. The complex voltages are then streamed to computing facilities at the Max-Planck-Institut f\"ur Radioastronomie in Bonn, and at the Forschungszentrum J\"ulich, where the data are coherently dedispersed, and folded at the pulse period. The final pulsar observation covers a bandwidth of ${\sim}71.5$~MHz ($95.3$~MHz for DE601) at a central frequency of ${\sim}153.8$~MHz ($149.9$~MHz for DE601 -- these parameters vary due to networking constraints). Depending on the pulsar brightness, the duration of each observation spans from 1 (for bright pulsars) to about 3 hours (for MSPs and fainter pulsars).\\

In this article, we study the binary MSP~J0034$-$0534 \citep{bhl+94,kvh16}, that was observed every week by four of the German LOFAR stations, namely DE601, DE602, DE603, DE605 (see Table~\ref{tab:obs} for a summary of the observations and of the observing setup of each used station\footnote{The metadata page of the GLOW dataset can be found at: \url{https://www.physik.uni-bielefeld.de/~soslowski/LOFARSS/index2.php}.}). The reasons why we chose this source are mainly: 1) J0034$-$0534 is the MSP that yields the most precise DM measurements in the entire GLOW dataset\footnote{Based on two years of observations with DE601 \citep{don14}. Among both MSPs and long-period pulsars, the most precise source of the entire GLOW sample is J2219+4754 (with a spin period of about 0.54 seconds), with J0034$-$0534 coming immediately after.}, 2) its ecliptic latitude of $-8.53$ degrees ensures a close approach to the Sun. PSR~J0034$-$0534 is also included in the European and International PTA (EPTA and IPTA, respectively) observing programmes \citep{dcl16,vlh16}.\\

\begin{table*}
\centering
\caption{Summary of the observations for PSR~J0034-0534. The channel width is 0.195~MHz for all the stations. The columns report, respectively, the station ID, timespan of the datasets, central frequency, bandwidth, number of original frequency channels, average timespan of an individual observation and number of observations.}
\label{tab:obs}
\begin{tabular}{ccccccc} 
	\hline
	Station & Timespan & $\rm f_{c}$ [MHz] & Bandwidth [MHz] & No of channels & $<T_{\rm obs}>$ [hr] & N. obs\\
	\hline
	DE601 & 2013-08-21 to 2018-02-17 & 149.9 & 95.31 & 488 & 2.1 & 194\\
	      & & & & & &\\
	
	DE602 & 2015-01-31 to 2018-02-17 & 153.8 & 71.48 & 366 & 2.5 & 142\\
	& & & & & &\\
	
	DE603 & 2014-02-12 to 2015-02-01 & 149.9 & 95.31 & 488 & 2.4 & 43\\
	      & 2015-02-07 to 2018-02-17 & 153.8 & 71.48 & 366 & & 140\\
	      & & & & & \\
	DE605 & 2014-03-07 to 2015-01-23 & 149.9 & 95.31 & 488 & 2.0 & 35\\
	      & 2015-02-06 to 2018-02-11 & 153.8 & 71.48 & 366 & & 146 \\
	\hline
\end{tabular}
\end{table*}

%

\section{Data analysis}\label{sec:datared}

\subsection{Timing and DM measurements}

\begin{figure*}
  \includegraphics[scale=0.75]{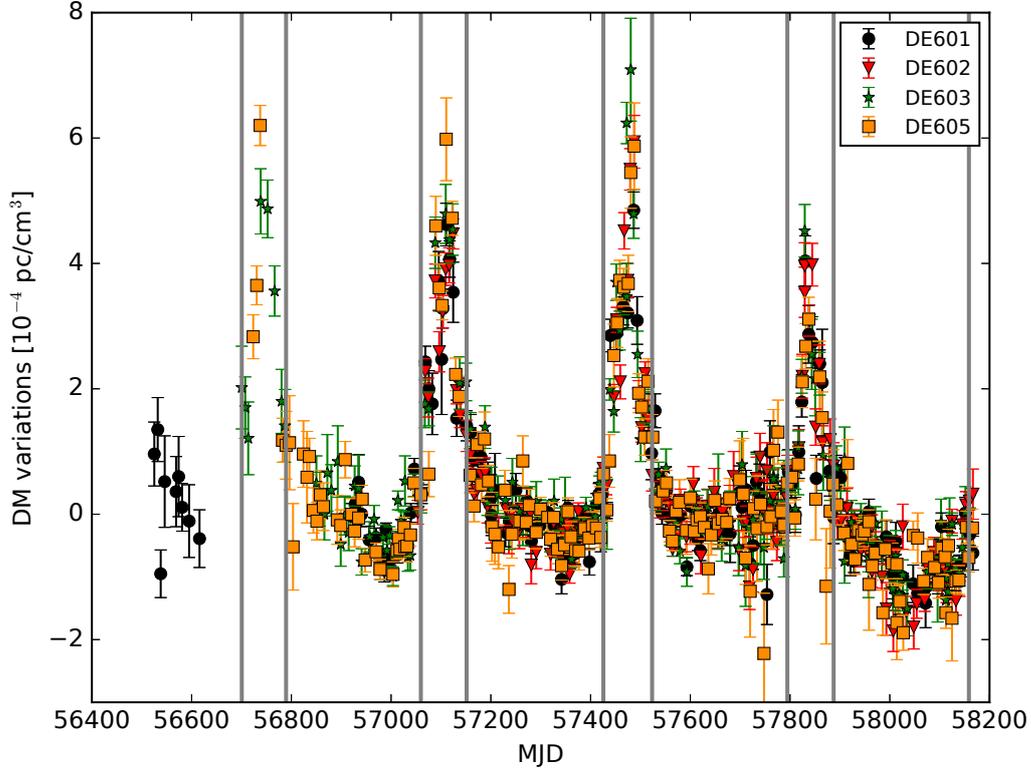}
  \caption{DM time series of PSR~J0034$-$0534, showing only the measurements with error bars smaller than $10^{-4}$~pc~~cm$^{-3}$. The different colours refer to different stations, while the vertical lines indicate a Solar angle of 50~degrees.}
   \label{fig:dmtimeseries}
\end{figure*}

The first step in the data analysis was radio-frequency interference excision. This was done through routines based on the \textsc{PSRchive} \citep{hvm04} and \textsc{CoastGuard} \citep{lkg16} software suites on ${\sim}23.8$~MHz-wide subbands independently.\\
An updated timing model was derived by using the T2 software package, starting from the IPTA ephemeris for PSR~J0034$-$0534 \citep{vlh16} and the DE601 dataset (for a description of the timing procedure, see e.g. \citealt{lk04}). For this purpose, the observations taken from DE601 were individually integrated over their observation durations and the number of frequency channels averaged down by a factor of 12, and a standard 2-D (frequency and phase) template was formed by summing across the entire DE601 dataset. Frequency-resolved times-of-arrival (ToAs) were obtained by cross-correlating the total intensity in each channel of identically averaged observations from all stations with that of the channels of the template \citep{tay92}. The template matching was carried out in the Fourier domain with a Markov-chain Monte-Carlo (MCMC) approach for improved estimates of the cross-correlation errors. 
 The IPTA timing model was first adjusted by removing any parameters related to red noise or DM variations, and the mean DM was estimated by fitting against the DE601 ToAs only. \\
This new timing model with the updated DM value was installed in fully time- and frequency-resolved observations of the datasets for each station, which were then calibrated by following the procedure described in \citet{nsk15}, based on the {\sc mscorpol}\footnote{\url{https://github.com/2baOrNot2ba/mscorpol}} software suite. For consistency, for all observations we only selected the part of the bandwidth that is common throughout the datasets, $\sim$71.5~MHz centred at 153.8~MHz, divided into 366 channels. Six channels from the edge of the bandwidth were additionally discarded, to obtain a final bandwidth of $\sim$70.3~MHz.\\  
We then derived the DM time series of the four datasets. We again chose the DE601 dataset to create the template, and we used the same averaging scheme described earlier (full time-averaging and a factor of 12 in frequency-averaging) for both the template and the observations. Through cross-correlation in total intensity in the Fourier domain with a MCMC approach, we generated a set of 30 frequency-resolved ToAs per observation, each ToA referring to a frequency channel of ${\sim}2.3$~MHz bandwidth. Outliers in the ToAs were identified as data lying further than three times the median absolute deviation from a robust fit for $f^{-2}$ (with $f$ being the frequency) performed in frequency on the timing residuals through Huber regression \citep{hub64,wks17}, and then eliminated. The ToA set of each observation was then fitted individually for DM by using the T2 software package. It is important to note that during the timing procedure, we did not employ any SW model (described in Section~\ref{sec:t2model}) operated automatically by T2. The final DM time series from all the stations are shown in Figure~\ref{fig:dmtimeseries}. The DM measurement uncertainties do not differ significantly between DE601 and the other stations, and this implies that the used template contained enough data to mitigate the risk of self-standardizing (see Appendix A of \citealt{hbo05a}). \\

\subsection{Disentangling IISM and SW}\label{sec:disentangling}

\begin{figure*}
   \centering
\begin{tabular}{cc}
\includegraphics[width=\columnwidth]{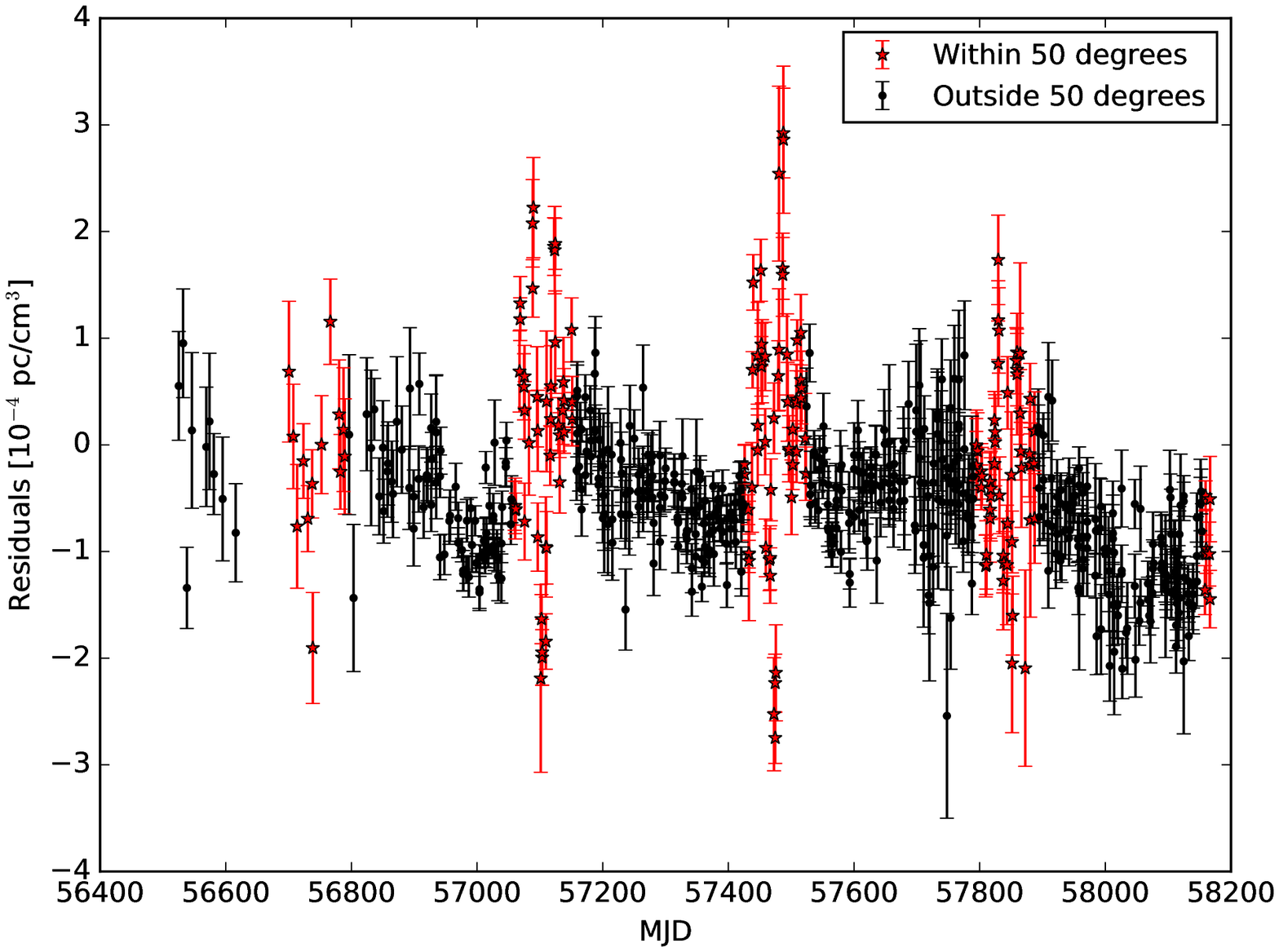}&
\includegraphics[width=\columnwidth]{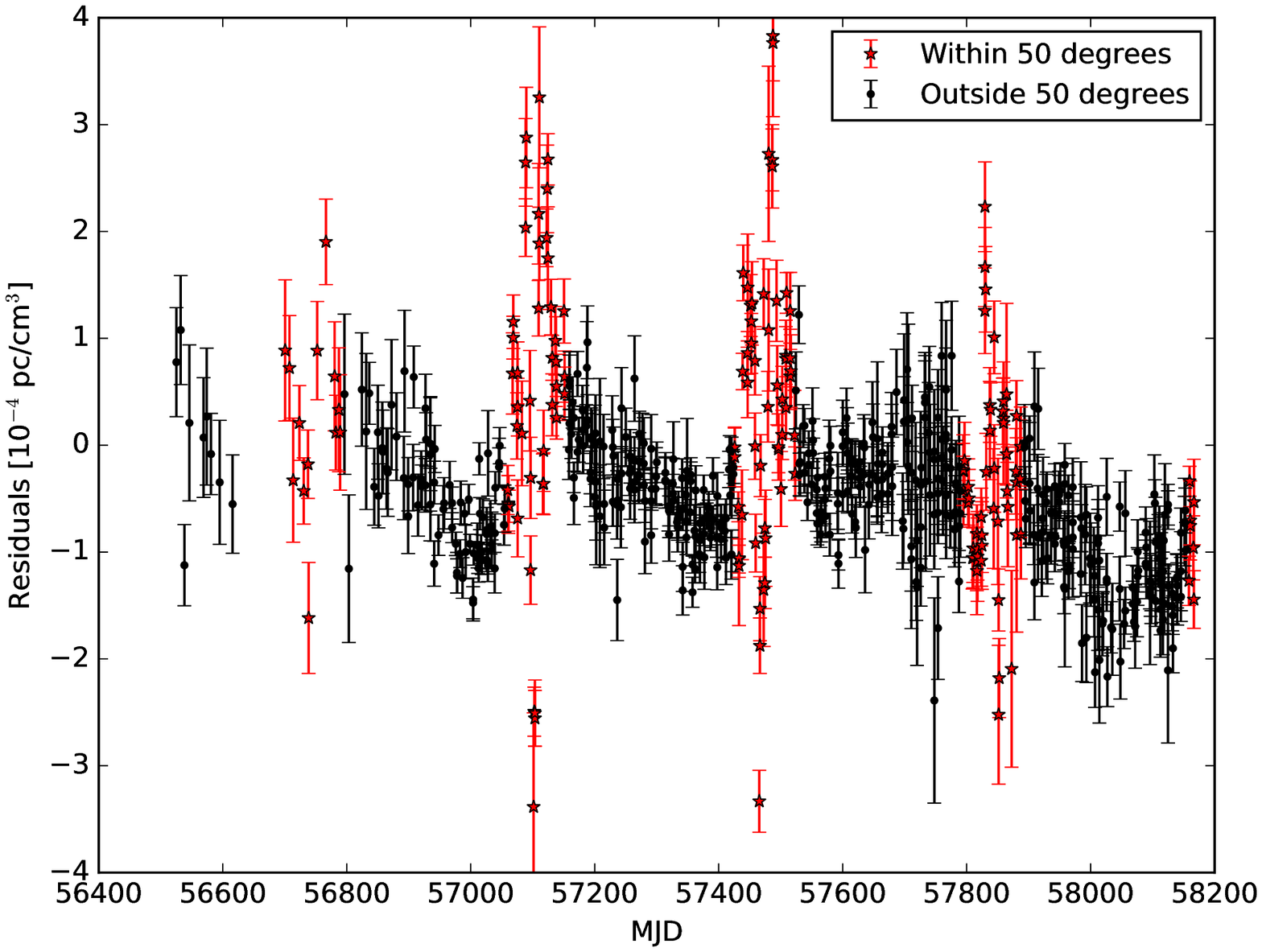}\\
\end{tabular}
\caption{Residuals between the time-variable T2 (left) and Y07 (right) models and the data. The DM residuals at Solar angles larger and smaller than 50~degrees are shown with dots and stars, respectively.}
   \label{fig:tenthswsubtraction}
\end{figure*}

The dispersive effects of the SW and the IISM are entangled, and it is necessary to subtract the interstellar signature from the DM time series to properly study the SW. \\
Usually, it would be safe to assume that a substantial fraction of the time series of DM variations centred around the anti-solar direction is fully dominated by the IISM. However, Figure~\ref{fig:dmtimeseries} shows that this is clearly not the case for our data on PSR~J0034$-$0534, where it is impossible to confidently state what timespans are dominated by the IISM, as significant variations are present throughout the GLOW dataset. For this reason, we used an iterative approach to separate the components.\\
First of all, we chose a suitable initial starting point for the SW contribution. As the aim was to test the effectiveness of the T2 and Y07 models, we separately used them as initial guesses (although the next sections show that they are both suboptimal representations of the SW effects) to extract two different estimates of the IISM. \\
We started the component separation by performing a year-wise linear fit of the T2- and Y07-model amplitudes to the DM time series, and we subtracted these fits from the data. As the Solar cycle has a duration of 11 years, it is reasonable to expect that there are appreciable yearly changes in the free electron content of the Solar wind.\\
The DM residuals, especially near the Solar approach, showed short-term structures that indicated clear discrepancies between the models and the data (see Figure~\ref{fig:tenthswsubtraction}).

\begin{figure*}
   \centering
\begin{tabular}{cc}
\includegraphics[width=\columnwidth]{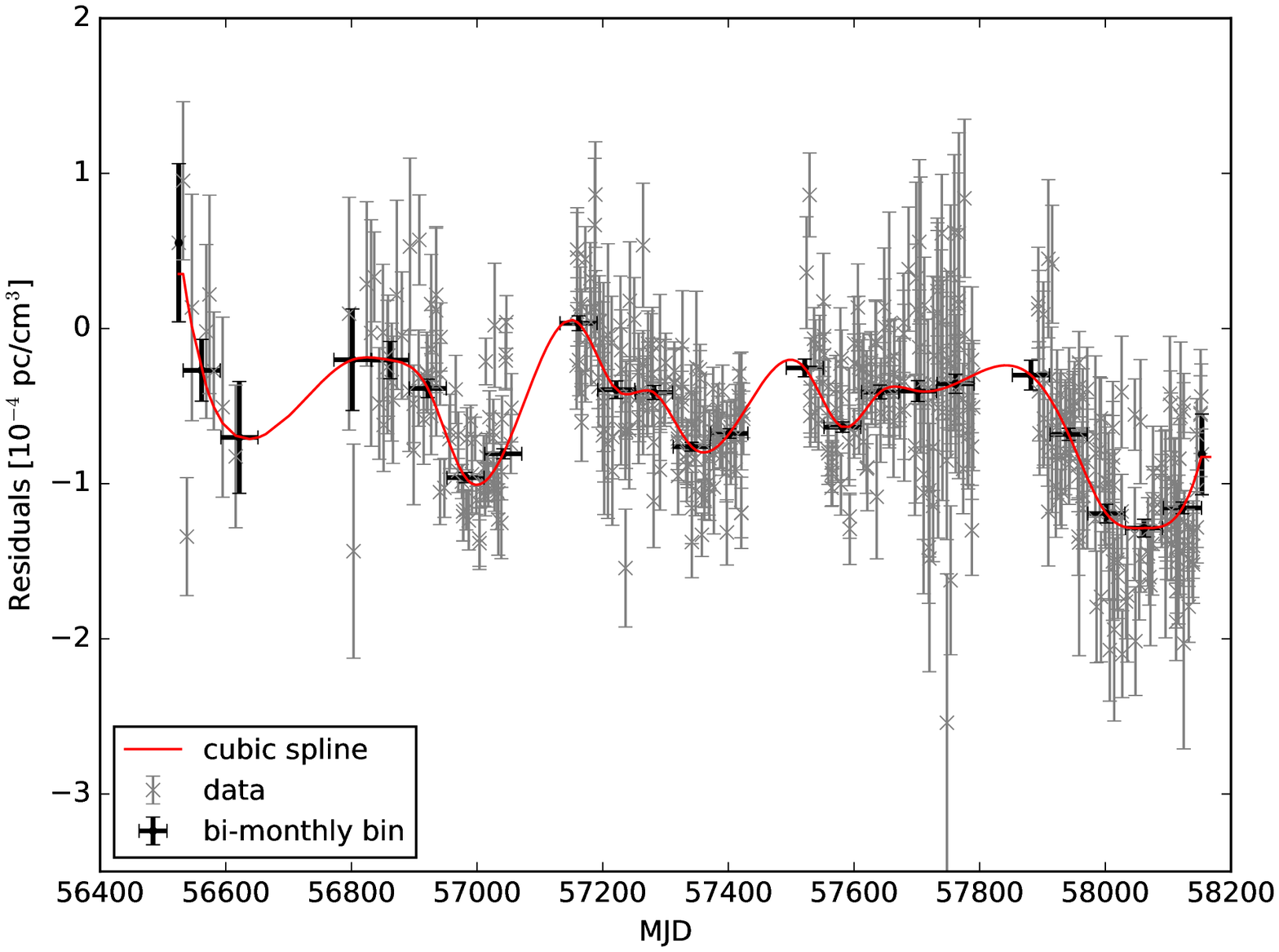}&
\includegraphics[width=\columnwidth]{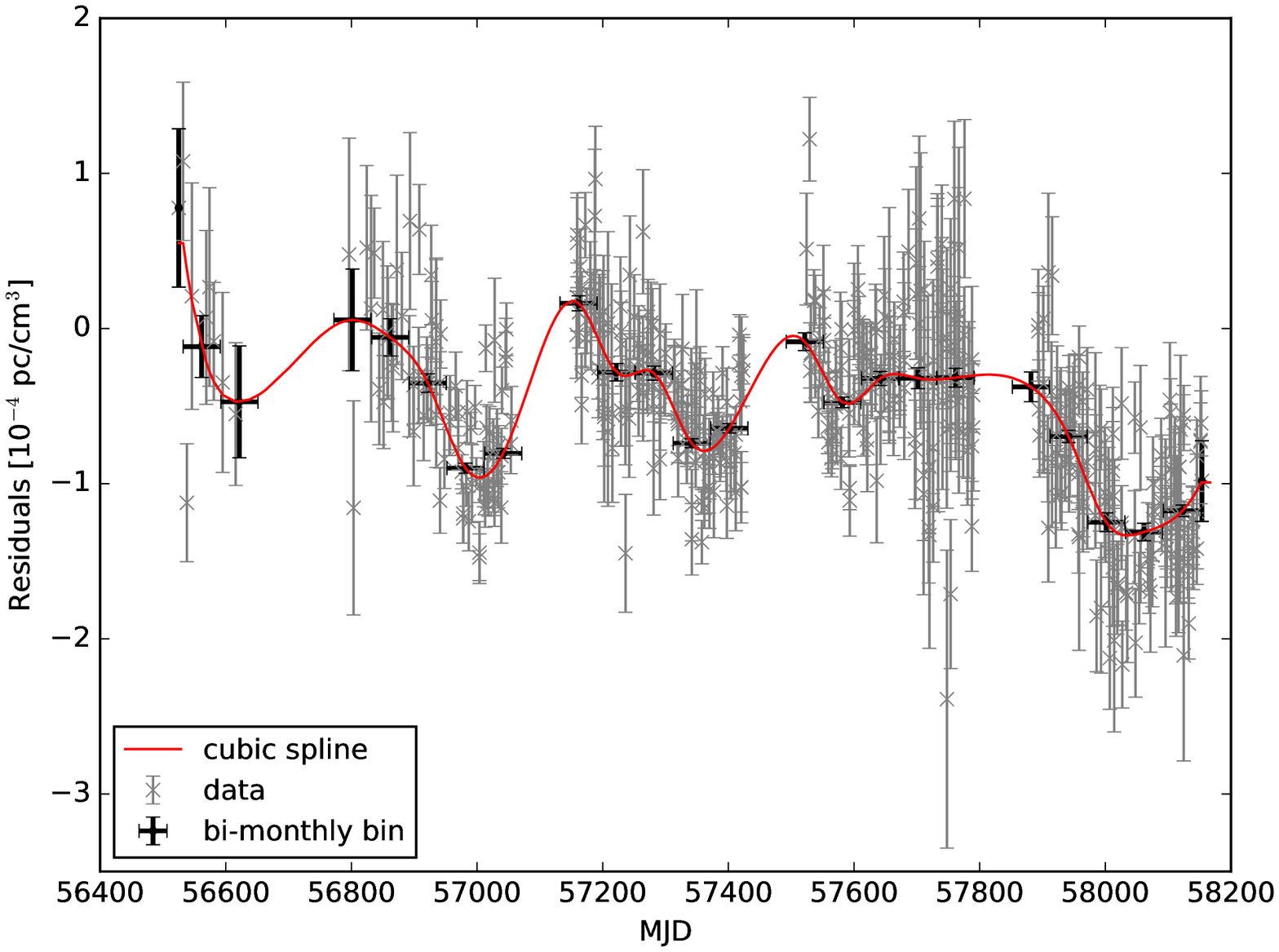}\\
\end{tabular}
\caption{Final residuals of the DM time series (in grey) after the iterative procedure to subtract the T2 (left) and Y07 (right) SW models. The black points show the bimonthly binning and the red line shows the fit for the cubic spline.}
   \label{fig:fifthspline}
\end{figure*}

To separate the contributions of the SW and IISM, we proceeded by removing (``windowing'') from the residual time series observations closer than 50~degrees to the Sun (see Section~\ref{sec:windowsize} for the determination of the windowing size). The remaining DM residuals were binned on bimonthly intervals, and then fitted with a cubic spline. With this procedure, we assumed that the spline mostly represents the IISM contribution. To obtain a more robust result, we iterated the described process by subtracting the spline from the original DM time series, and then repeating the year-wise fitting of the T2- and Y07-model amplitudes to the DM residuals. This operation was repeated ten times, to reach a convergence in the fit for the SW model (the fits for the cubic spline at the tenth iteration are shown in Figure~\ref{fig:fifthspline}). \\

 \begin{figure*}
    \centering
 \begin{tabular}{c}
\includegraphics[scale=0.6]{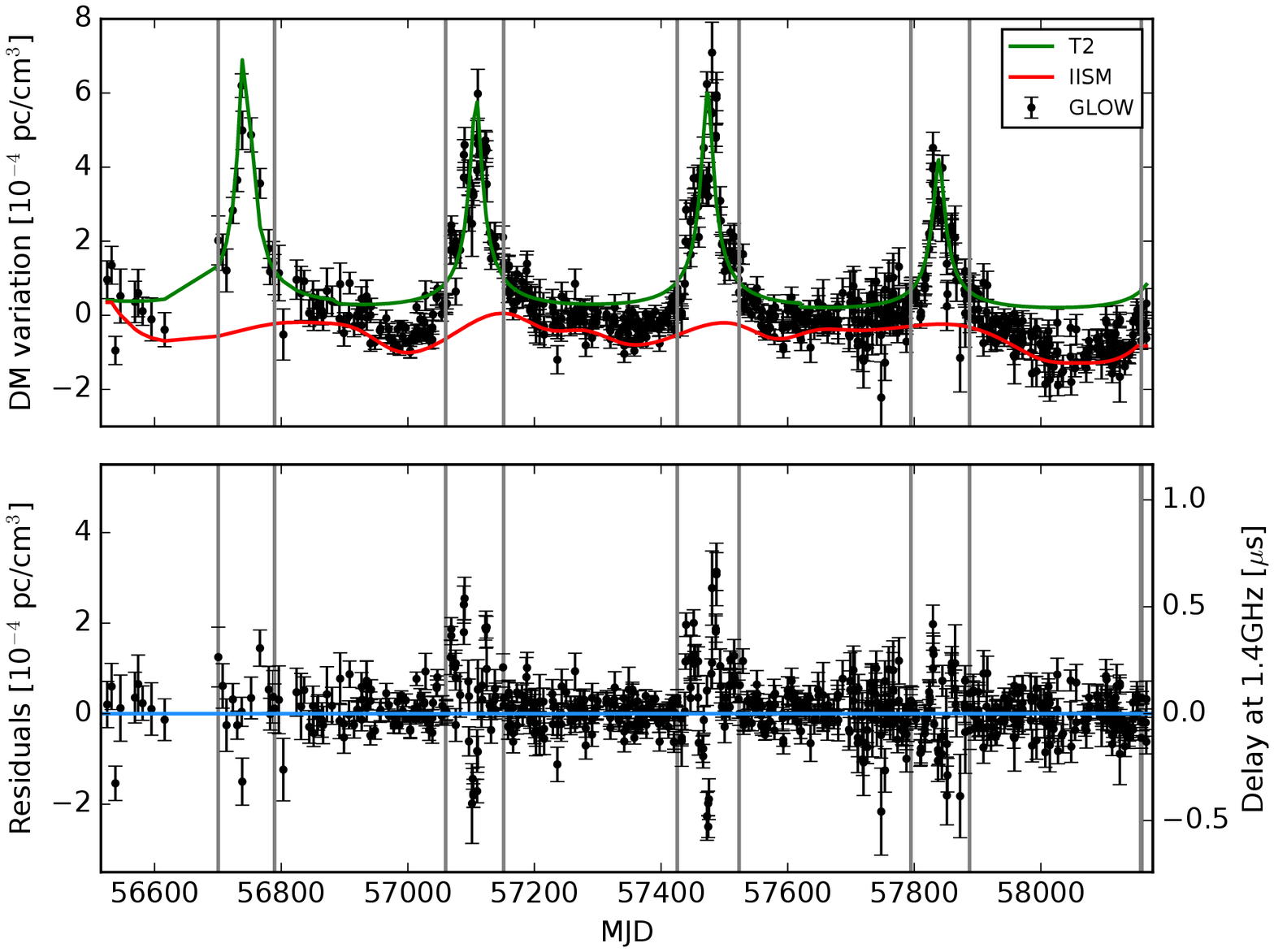}\\
 \includegraphics[scale=0.6]{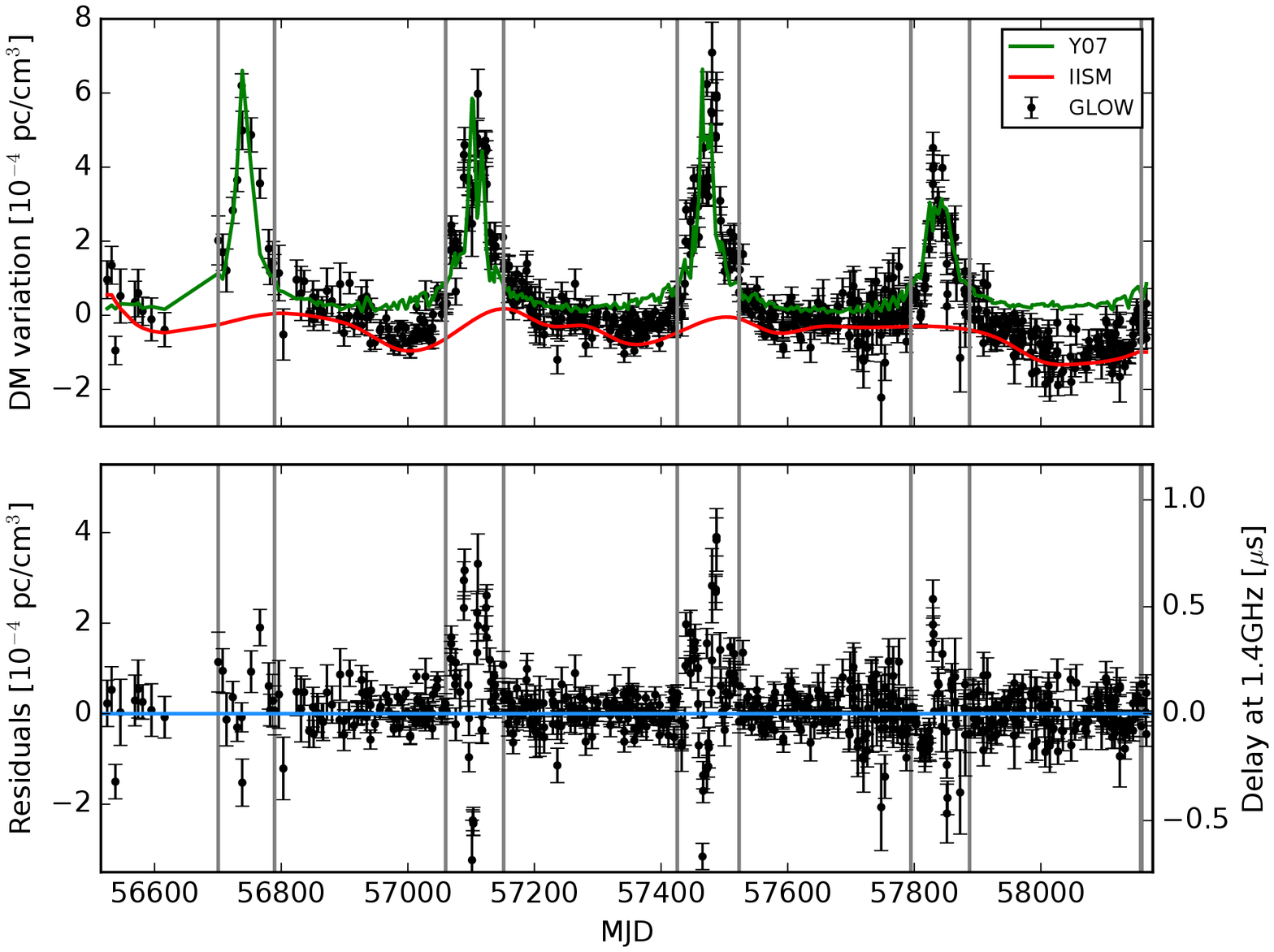}\\
 \end{tabular}
  \caption{\textit{Upper panels} -- DM time series estimated from the GLOW data (in black), estimates of the IISM through the spline fit (in red), and SW contribution to the DM as estimated by the T2 model (in green, \textit{upper plots}) and by the Y07 model (in green, \textit{lower plots}). \textit{Bottom panel, left y-scale} -- DM residuals obtained by subtracting the IISM and SW modelled contributions from the GLOW data. The blue line marks the 0 of the y-axis. \textit{Bottom panel, right y-scale} -- As the left y-scale, but showing the corresponding time delays at 1.4~GHz. The grey vertical lines indicate a Solar angle of 50 degrees.}\label{fig:y07t2performance}
 \end{figure*}

The global results of the IISM disentangling for both of the SW models are shown in Figure~\ref{fig:y07t2performance}. We can think to the DM time series shown in Figure~\ref{fig:dmtimeseries} as:
\begin{equation}
 {\rm DM}(t) = {\rm SWM}(t) + {\rm IISM}(t) + {\rm R}(t)
\end{equation}

Where $\rm SWM$ is the contribution as computed by the SW model (in green in upper panels), $\rm IISM$ is the interstellar contribution (red) as computed through the spline fitting method and $\rm R$ are the residuals (bottom panels). Both of the SW models were improved by retaining the yearly amplitude fit (to account for the Solar variability). We stress that the implementation of a variable amplitude was already carried out by \citealt{ych12} (follow-up article to \citealt{yhc07}) to better interpret the analysis of the SW magnetic field, and, as mentioned in Section~\ref{sec:t2model}, the current implementation of the spherical model in T2 also allow an arbitrary amplitude. \\

\begin{figure*}
 \centering
 \includegraphics[scale=0.65]{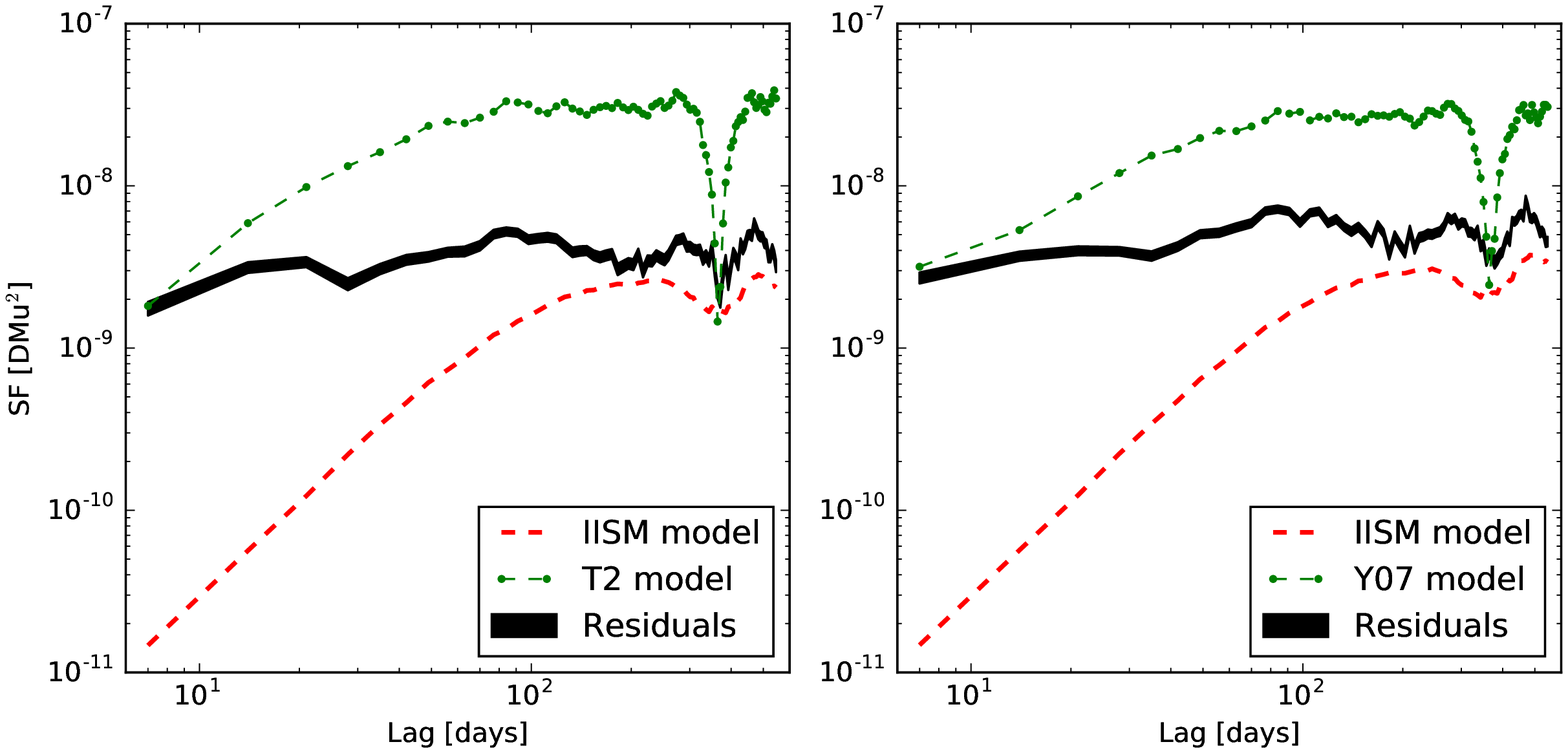}
 \caption{SFs of the various components of Eq.~\ref{eq:components} for the T2 (left) and Y07 (right) SW models. The colour scheme is the same used in Figure~\ref{fig:y07t2performance}: the green, dash-dotted lines represent the SFs of the SW models, the dashed, green lines are the SFs of the estimated IISM approximations, while the black shaded areas are the SFs of the residuals.}\label{fig:sfs}
\end{figure*}

To evaluate the results of our IISM disentangling method, we computed the structure functions (SF) of the spline model (IISM), of the SW model and of the residuals. The SF is defined as:
\begin{equation}\label{eq:components}
 D(\tau) = \langle ({\rm DM}(t) - {\rm DM}(t+\tau) )^2 \rangle,
\end{equation}

where $\tau$ is the time lag that separates the two DM values. The SF shows the correlations among different time lags, and it can be used to investigate the amount and type of plasma turbulence that affected the propagation of pulsar radiation. Because of the irregular sampling of the observed DM time series, the SFs were computed over time lags at multiples of 7 days. Also, we did not compute SFs for lags larger than a third of the observing time span, since the information carried by all the measurement pairs at larger lags becomes increasingly statistically similar. The results for both of the SW models are shown in Figure~\ref{fig:sfs}.\\
When computing the SF of the residuals, we obtained an estimate of the errors due to the uncertainties in the DM values in each of the SF bins through a Monte Carlo procedure, by simulating a thousand new DM time series based on the original one, with each measurement being extracted from a Gaussian distribution with mean and standard deviation equal to the original DM measurement and its uncertainty, respectively. Again, only for the residual SF we estimated the white noise level as the SF values at $\tau<3.5$~days, half the width of the first bin, and we subtracted this estimate from the SF. \\

A few points are immediately clear from the computed SFs. First of all, the IISM component lies at a lower level with respect to the other two components. It gets closer to the SW model SFs only at the largest lags, indicating that the IISM contains only a low-frequency signal\footnote{Note though that, by construction, the spline approximation acts as a low-pass filter. Besides this, the IPTA data release \citep{vlh16} shows that this pulsar has a significant DM1 and DM2 parameters, that would show up as a parabolic and a quartic trend, respectively, in the SF, see e.g. \citealt{lcc16}. However, both of the DM derivatives are extremely small with respect to the range spanned by the IPTA values, and hence it is possible that the two trends are indistinguishable in the SF of the IISM approximation.}. The magnitude of the IISM approximation SF implies that the IISM is relatively constant in time across the examined time span. In addition to this, the IISM approximation clearly absorbs part of the SW effects as it can be seen from the IISM SF dip at $\tau\sim1$~year. This artificially lowers the residual values, that hence only represent a lower limit on the real residuals. The residual SFs appears largely representative of a white noise time series, except for the signature at a periodicity at $\tau\sim1$~year. This periodicity, as can be clearly seen from Figure~\ref{fig:y07t2performance}, arises from the DM structures computed during the Solar transit, that both of the SW models fail to reproduce (and by construction, were not modelled by the spline approximation of the IISM). The SF of the SW models lies above the other two, indicating that the SW is the component that shows the greater magnitude of time variations in the examined time span.

\section{Quantitative evaluation of the two Solar-wind models in pulsar timing}\label{sec:evaluation}

Referring to Eq.~\ref{eq:components}, we now proceed with analysing the residuals, ${\rm R}(t)$ from Eq~\ref{eq:components}, for the two SW models. Theoretically, the residuals for both of the models should consist of pure white noise measurements only (essentially a stationary, stochastic statistics) if all the modelling was successful. If they are not, then they will represent a lower limit on the unmodelled SW contribution, as the IISM signal is marginal in terms of amplitude. \\
 
While the two models appear to be able to correct the Solar effects at large angular distances from the Sun (which is probably partly due to our IISM component absorbing the slower variations on timescales of months to years), they both fail to account for the SW influence at the Solar approach. It is interesting to note that the predictions for the SW contribution in the anti-solar direction are, on average, higher than  $10^{-5}$~pc/cm$^{3}$ for both of the models. Because the uncertainties on our DM measurements for PSR~J0034$-$0534 are typically less than $10^{-4}$~pc/cm$^{3}$, the SW is always an important contribution in this dataset -- even at large Solar angles. \\

\begin{figure*}
\centering
\includegraphics[scale=0.6]{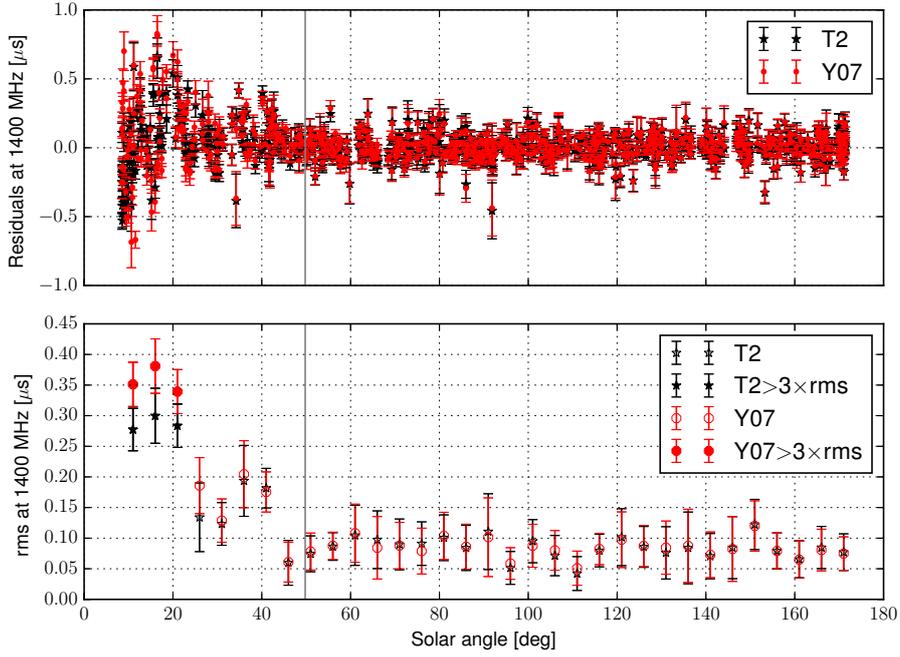}
\caption{\textit{Upper panel} - Time-delay residuals between the data and the Y07 and T2 models at 1400~MHz (represented as circles and stars respectively) versus the Solar angle. The grey vertical line corresponds to a value of 50~degrees for the Solar angle. \textit{Bottom panel} - rms of the time-delay residuals shown in the upper panel, binned every 5~degrees. The bins where the rms exceeds three times the rms computed at Solar angles larger than 50~degrees are indicated with filled markers.}
   \label{fig:rms}
\end{figure*}

We proceeded by converting the DM residuals in the bottom panels of Figure~\ref{fig:y07t2performance} to time delays at 1400~MHz, the preferential observing frequency in PTA experiments \citep{dcl16,rhc16,vlh16,abb18}. The top panel of Figure~\ref{fig:rms} shows the time-delay residuals as a function of the Solar angle. After that, we binned the time-delay residuals for every 5 degrees in Solar angle, and computed the root-mean-square (rms) for each bin. We also calculated the rms over all the data at Solar angles larger than 50~degrees: such ``off-Sun'' rms can be considered as the reference noise level reachable by an optimal modelling of both the IISM and SW contributions, and as an estimate of the standard deviation of the time-delay residuals that effectively quantifies the sensitivity of our dataset. We then identified the bins whose rms exceeds three times the standard deviation.\\

\begin{figure*}
\centering
\includegraphics[scale=0.6]{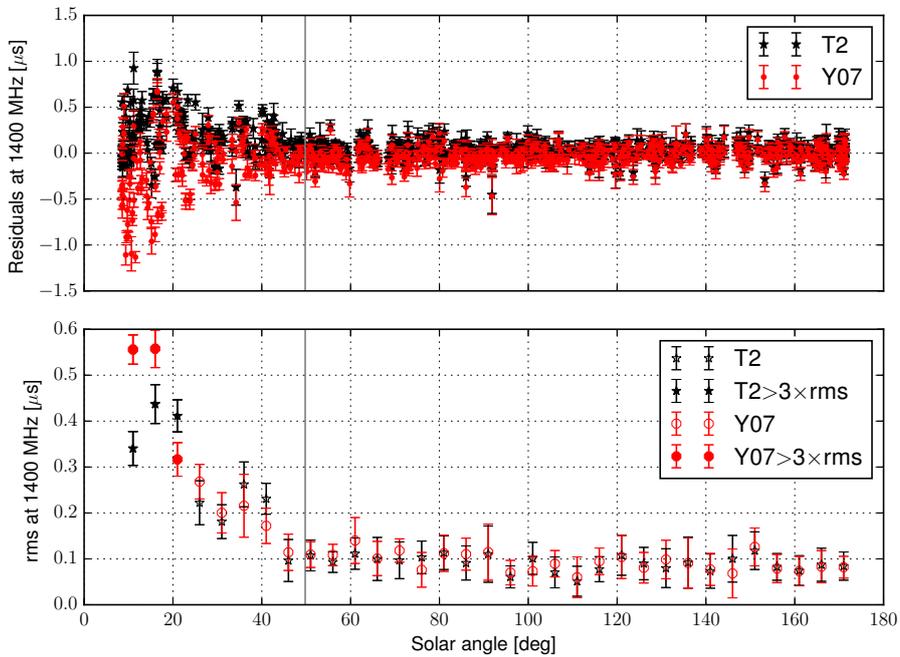}
\caption{Same as in Figure~\ref{fig:rms}, but without variable amplitude fit in any of the used models.}
   \label{fig:rms_nkf}
\end{figure*}

Those results are shown in the bottom panel of Figure~\ref{fig:rms} and they indicate that the performance of the two models is remarkably similar. Between $\sim$20 and $\sim$40~degrees from the Sun, both models yield time-delay residuals whose rms is within three times the threshold computed around the anti-solar direction. Closer than that, neither of the models is able to correctly predict the SW dispersive behaviour. However, the plot shows clearly that the T2 model performs better at almost all the angular distances within 50 degrees, reducing the excess rms by 2 to $\sim$28\%.\\
This is in contrast with the findings of \citet{yhc07}, where the authors claim that the Y07 model is superior to the spherical approximation in correcting the timing residuals. There are three possible reasons at the basis of their conclusions: (1) the relatively lower precision that DM measurements provide at the higher observing frequencies used by \citealt{yhc07}, (2) a difference in the performance of the Y07 model depending on the helio-latitude (since they studied a different pulsar), and (3) the lack of a meaningful year-wise variable fit of the model amplitude. However, as shown Figure~\ref{fig:rms_nkf}, this last potential cause is likely to be the least significant. Figure~\ref{fig:rms_nkf} was obtained by performing the same steps as described in Section~\ref{sec:disentangling}, but without applying the year-wise variable scaling factor in the final iteration of the process. We can infer from this plot that our conclusions do not change significantly, since only at 21~degrees of Solar angle does the Y07 model clearly outperform the T2 approximation, by $\sim$30\%. Moreover, the results that drove the conclusions of \citet{yhc07} were obtained at closer angular distances than $\sim$20~degrees. As a matter of fact, \citet{yhc07} mostly relied on the datasets collected at the Nan\c{c}ay Radio Telescope during four solar passages of PSR~J1824$-$2452A (with an ecliptic latitude of $-$1.55~degrees), and first presented by \citet{cbl+96}. The right-hand panel of Figure 3 in \citet{yhc07} shows that at epochs corresponding to 20 to 40 degrees from the Sun, the T2 and Y07 models perform similarly, while the main differences happen at closer Solar angles. Thus, either the DM precision or a helio-latitude dependence are the most likely explanations of the discrepancies with respect to our results. Note that in the EPTA data release \citep{dcl16}, the ToA rms of J0034$-$0534 is 4~$\mu$s (based on datasets collected with the Nan\c{c}ay Radio Telescope at L-band and the Westerbork Synthesis Radio Telescope at three different frequencies from P- to L-band). This is the reason why the noise introduced by the unmodelled parts of the SW was undetected in the current EPTA dataset.

\subsection{Causes for the inconsistencies}

We now discuss why neither the T2, nor the Y07 model are able to properly predict the SW effects. \\

In the case of the T2 model, the main issue is that it does not attempt any correction for the high-frequency temporal fluctuations of the electron density in the SW. While the T2 model appears to be able to capture the long-term trend of the SW (if and when a yearly-variable amplitude is applied), this impression is biased by absorption of some of the long-term signals into our IISM approximation. Another marginal issue is that it is a purely radial model (as well as the Y07), and hence the ingress (when the source approaches the Sun) and the egress (when the distance between the source and the Sun increases) of a pulsar are modelled in the same way. In reality, the electron density of the SW is not constant with respect to the radial distance to the Sun (although this approximation is usually done), and it is not constant in time. \\

On the other hand, the Y07 model was indeed designed to track the rapid variations of the SW. However, if such variations are not optimally modelled, the overall effect is to introduce an additional quantity of noise. An inaccurate prediction of the SW fluctuations can happen for numerous reasons. (1) The synoptic maps of the Solar corona come from an extrapolation derived without any information about the polar field, and that relies on two strong assumptions: the existence of a current-free space with respect to the photosphere, and that the photosphere and the corona have the same angular velocity. This implies, e.g., that the time stamps of the coronal maps, or their latitude labelling, might not be correct but marginally offset. (2) The assignment of slow or fast wind to each segment of the projected LoS does not take into account the presence of CMEs, CIRs, streamers and gaps in the SW, that can increase or decrease the DM with respect to the model predictions. (3) The SW speed, and the thickness in size of the slow wind belt are considered time-independent, which does not reflect their known behaviour. Moreover, the SW speed should not be identical for both slow and fast wind, instead they are typically unique and depend on the overall Solar activity. As an additional note, however, the Y07 model may achieve significant improvements by testing different scaling laws to describe $\rm n_{e,slow}$ and $\rm n_{e,fast}$, by introducing a dependency on the heliographic latitude and by applying separate, time-variable amplitudes for the two SW phases (while, currently, the time-dependent amplitude that we introduced in Section~\ref{sec:disentangling} was the same for both of the SW phases). \\

\begin{figure*}
\centering
\includegraphics[scale=0.6]{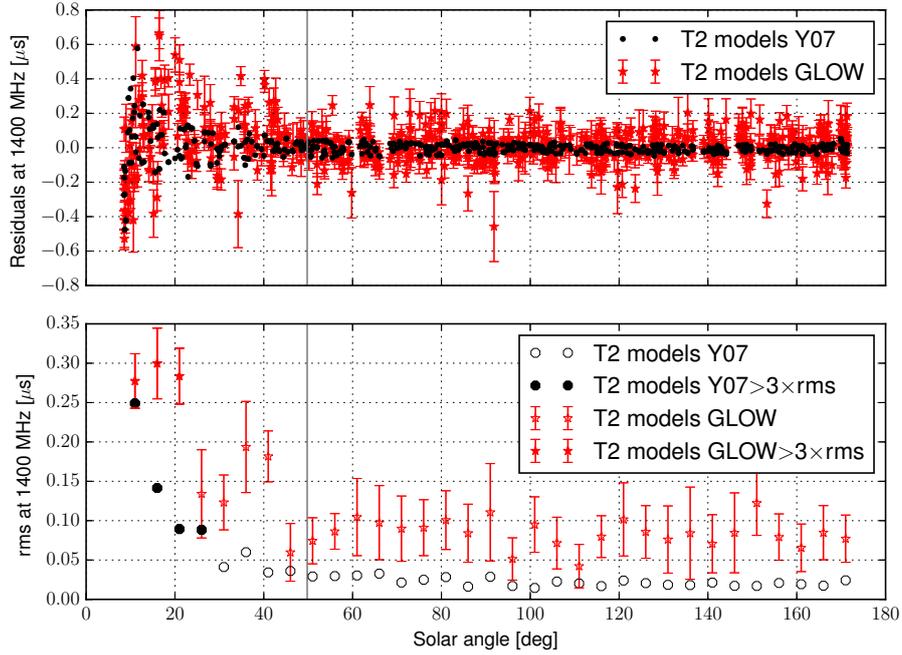}
\caption{Same as in Figure~\ref{fig:rms}, but using the same model (T2) on different ``datasets'' (Y07, in black, and GLOW, in red).}
   \label{fig:rms_y07}
\end{figure*}

We performed a test to verify if the amount of high-frequency, temporal fluctuations in the time series of the DM variations predicted by the Y07 model can be considered a practical representation of the real DM variations introduced by the Sun (i.e., the ones that we see in the dataset). For a proper comparison, we need to subtract the long-term trend of the SW from the Y07 model, and both the long-term trend and the IISM contribution from the real data. For this purpose, we applied the T2 model to the DM time series as computed by Y07, to subtract the long-term trend of the SW, and we subtracted both the T2 model and the IISM contribution from the real data (i.e., resulting in the black-coloured data of Figure~\ref{fig:rms}). Figure~\ref{fig:rms_y07} reports the results, and shows that the Y07 model still underestimates the amplitude of high-frequency temporal features that characterise the real data, however, none of the discrepancies goes beyond one order of magnitude, and usually remains within a factor 3 or 4. This implies that the Y07 model provides a reasonable order-of-magnitude estimate of the kind of variations to expect, but cannot be used as perfectly predictive.

\subsection{Test case: impact of CMEs}

\begin{figure}
\centering
\includegraphics[scale=0.45]{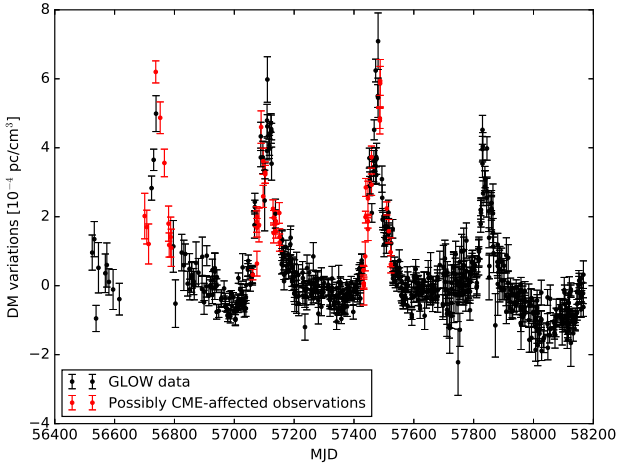}
\caption{Observations in the GLOW dataset for PSR~J0034$-$0534 potentially affected by CMEs (in red).}
   \label{fig:cme}
\end{figure}

\begin{figure*}
\centering
\includegraphics[scale=0.6]{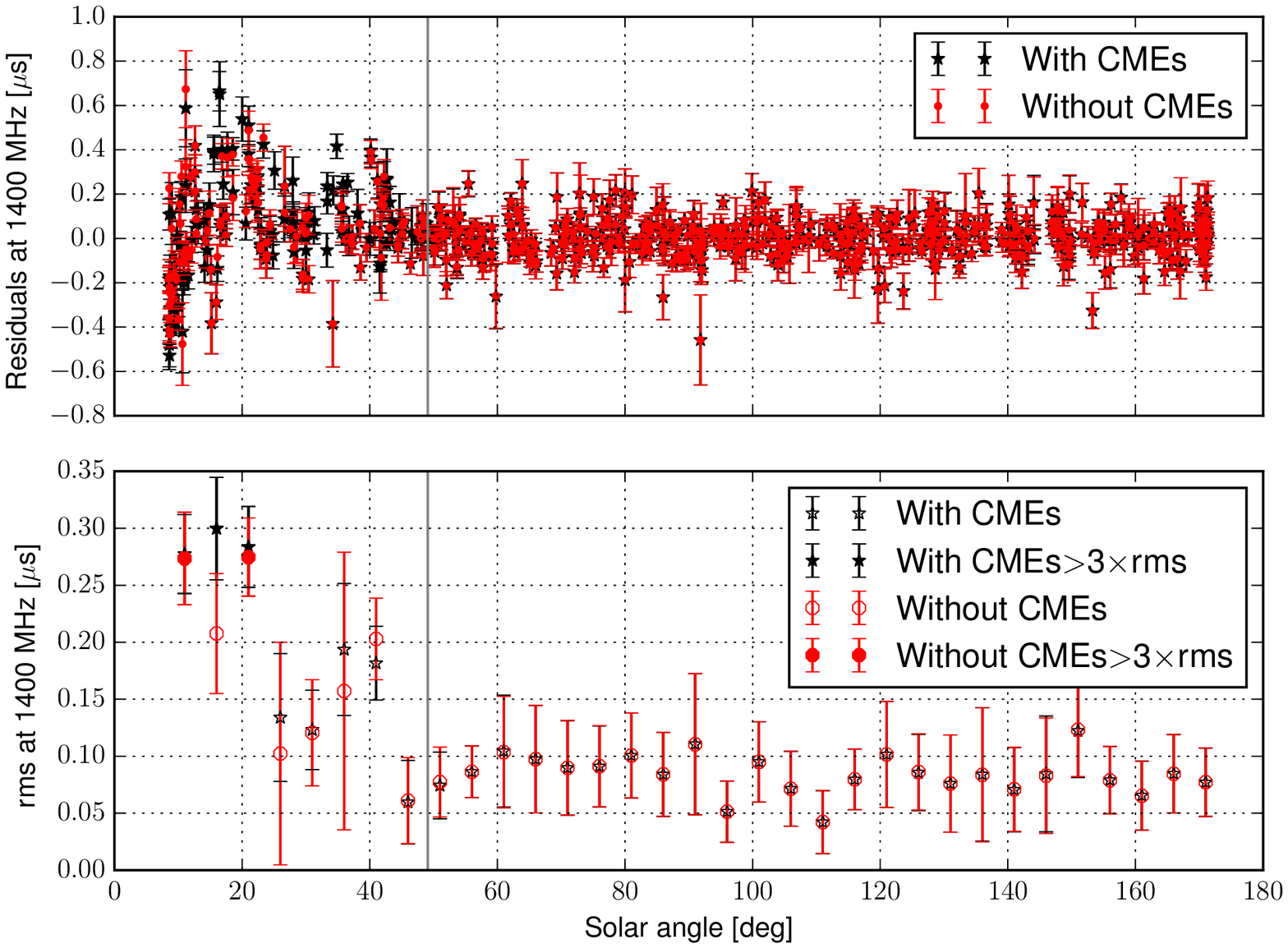}
\caption{Same as in Figure~\ref{fig:rms}, but comparing the effect of the T2 model on the complete GLOW dataset (in black) with respect to the GLOW dataset minus observations potentially affected by CMEs (in red). }
   \label{fig:rms_cme}
\end{figure*}

As mentioned in Sections~\ref{sec:intro} and \ref{sec:swstructure}, CMEs are masses of magnetised plasma that detach from the Solar corona above an active region -- hence, more CMEs are expected during the periods of maximum Solar activity. As CMEs are not included in either of the two SW models available in pulsar timing, we aim to infer if they can be used to explain the discrepancies shown by Figure~\ref{fig:rms}. \\
We obtained the list of CMEs emitted during the time interval spanned by our observations from the catalogue provided by the Large Angle and Spectrometric COronograph onboard of the SOlar and Heliospheric Observatory satellite (LASCO/SOHO)\footnote{\url{http://cdaw.gsfc.nasa.gov/CME_list/UNIVERSAL/text_ver/univ_all.txt} \citep{gym09}. Note that the catalogue lists detected CMEs only up to the end of October 2017, while our dataset extends until the 17th of February, 2018. However, this does not affect our analysis, as CMEs emitted far from the Solar approach (that happens between March and April) are not expected to have a significant impact on DM measurements.}. We then made a series of hypotheses. (1) We did not consider any CME whose linear speed was less than 400 km/s because of the SW drag \citep{ssc15}. (2) We accepted an error of 15~degrees on the position angle of the CME emission, making the Solar north coincide with the Ecliptic north, for the sake of simplification. (3) Because the majority of the discrepancies between the models and the data happens at close distances to the Sun, we limited the search only to observations whose Solar angle is less than 50~degrees. (4) We assumed that a CME travels along a straight path without acceleration.\\
For each CME in the list, we computed the travel time to reach an observation at a certain angular distance, and we performed a primary selection by collecting all the observations that were taken before the expected CME arrival time. We further refined the choice by selecting only those CMEs emitted in the same quadrant occupied by the pulsar on the helioprojective plane. A subgroup of CMEs are directly emitted toward Earth, and thus the second selection criterion (the source location) was not applied to them.\\
We flagged 52 observations as possibly affected by CMEs (see Figure~\ref{fig:cme}). While a more thorough search for CMEs is beyond the scope of the article, we stress that some of the assumptions made during the analysis are extremely conservative, and that the number of affected observations is likely much smaller.\\
Figure~\ref{fig:rms_cme} shows the change between the T2 modelling of the complete GLOW dataset, and the GLOW dataset minus the flagged observations. The most notable difference happens at Solar angles of about 16~degrees, where the rms seems to drop by 44\%. However, the number of observations that are used to compute the rms in that bin also drops, from 21 to 13 (this is the largest decrease among the three inner bins). Besides, while at almost all the angular distances there appears to be an improvement, the absolute rms values are still above the acceptable levels at small Solar angles, and do not drop below 200~ns. We thus conclude that CMEs alone cannot explain the differences we find in our observations, relative to the two SW models investigated here. \\

 \subsection{Testing different windowing sizes}\label{sec:windowsize}

\begin{figure*}
\centering
\includegraphics[scale=0.6]{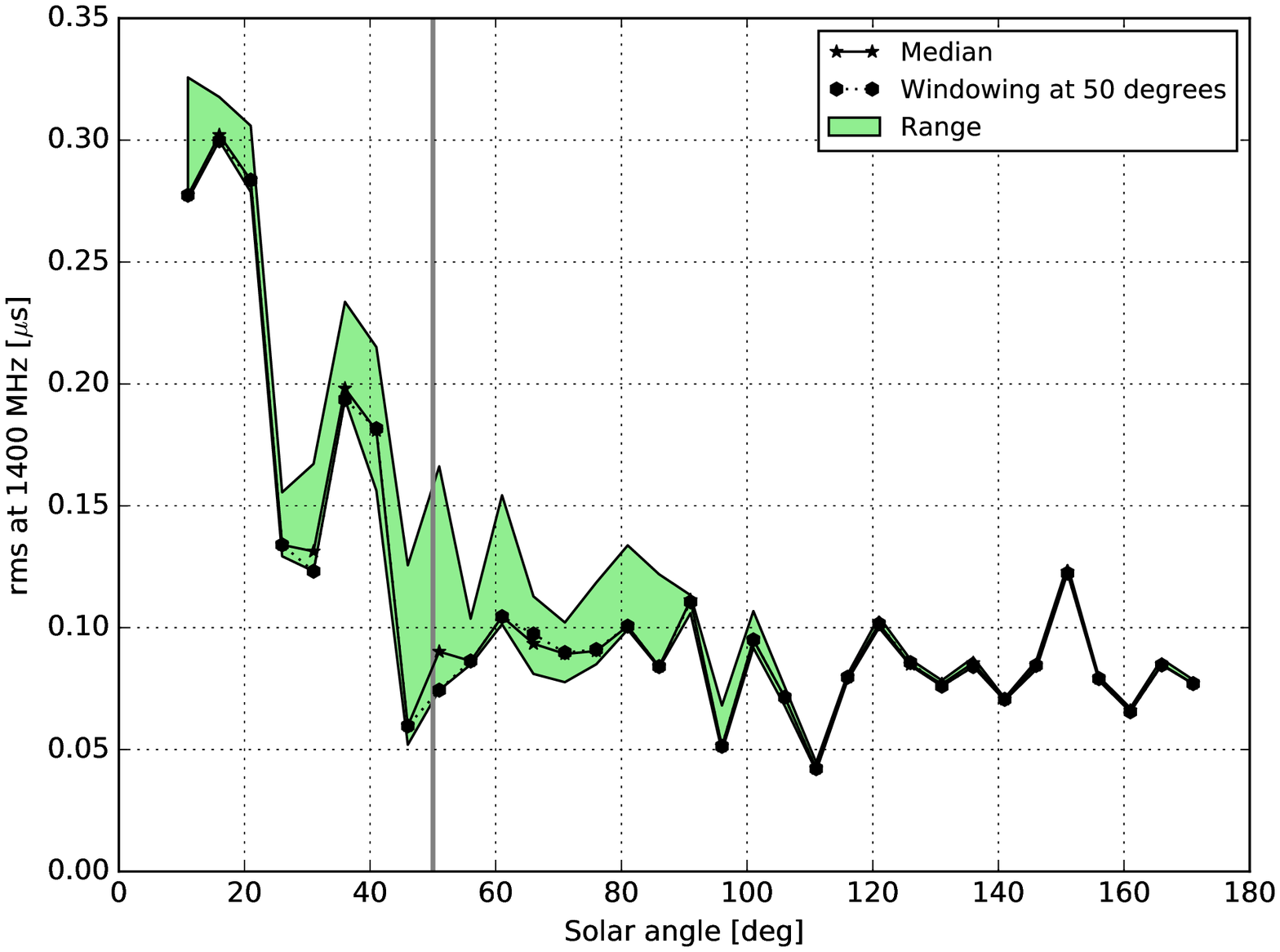}
\caption{RMS of the time-delay residuals obtained by subtracting the T2 model and the IISM contribution at 1400~MHz versus the Solar angle, evaluated for variable windowing sizes, as explained in Section~\ref{sec:evaluation}. The green shaded area is included between the upper and lower extremes of the residual timing rms achievable by choosing different windowing sizes. The median is shown as a continuous black line and with star-shaped markers, while the line obtained by applying a 50-degree window is shown by the dotted black line and the hexagonal markers. The grey vertical line marks a Solar-angle value of 50~degrees.}
   \label{fig:differentwindowings}
\end{figure*}

It is reasonable to question whether the 50-degree windowing that we applied while estimating the IISM contributions may have biased our results. We thus proceeded as follows. We reproduced the IISM-SW separation as described in Section~\ref{sec:disentangling} with a different series of window sizes, ranging from 30 to 80~degrees in steps of five degrees, and by using the T2 model with a variable amplitude. We then obtained the same kind of plots shown in the bottom panel of Figure~\ref{fig:rms} for each of these windows, and we identified upper and lower rms bounds for each of the curves, as well as the median values. While the upper limit (yielded by the widest windowing of 80~degrees) is likely to contain unmodelled IISM contributions, the majority of the curves corresponding to windows narrower than 65~degrees cluster around the median. Among them, the most distant curves from the median are the ones corresponding to the narrowest windows, which are likely to underestimate the contribution from the Sun (as the spline fit absorbs part of it). Among the remaining curves, the ones derived by applying a windowing size of 50~degrees (and 55, not shown here) are the closest to the median, see Figure~\ref{fig:differentwindowings}. Thus, we chose a windowing of 50~degrees as the most unbiased one.

\section{Implications for high-precision pulsar timing}\label{sec:ptaimpact}

 None of the mitigation routines presently available for pulsar timing perform well on low-frequency observations of a moderately low-ecliptic pulsar.\\ 
 Low-frequency observations clearly demonstrate the discrepancies between both of the T2 and Y07 approximations and the data, and Figures~\ref{fig:y07t2performance} and~\ref{fig:rms} highlight the effects of this inefficient modelling at 1400~MHz, that is usually the most common frequency for pulsar observations in high-precision timing experiments such as PTAs. In particular, Figure~\ref{fig:rms} shows that the rms of the time-delay residuals remains as high as 100~ns at Solar angles larger than 40~degrees, when a 100~ns rms level in the timing residuals is usually indicated as the rms magnitude to aim for in PTA data. It is worth stressing that the spline fitting described in Section~\ref{sec:disentangling} probably absorbed part of the Solar effects that the Y07 and T2 models were not able to subtract, and which appears as a yearly-periodic signature in the SF (as shown in Figure~\ref{fig:sfs}). Therefore, Figure~\ref{fig:rms} represents a best-case scenario, and should be interpreted as a lower limit on the rms introduced by the imperfect modelling of the SW.\\
  Individual PTA experiments have different approaches to SW mitigation. The IPTA (that embraces all the ``regional'' PTA projects) usually excludes ToAs closer than 5~degrees to the Sun \citep{vlh16}. \\ 
  The Australian PTA (the ``Parkes PTA'', \citealt{rhc16}) does not take into account ToAs within 10 days from the Solar approach for each of the pulsars in the array (Os\l{}owski, private communication). To date, this PTA project has presented the most stringent upper limit on the GWB: $10^{-15}$ in terms of characteristic amplitude \citep{srl15}. As the analysis 1) was performed on one individual pulsar observed at 3~GHz (J1909$-$3744) and with an ecliptic latitude of $\sim-$15, and 2) was not based on spatial correlation, but uniquely on the red noise level in the timing residuals, we consider it unlikely that the SW effect might have affected their analysis. Moreover, by extrapolating the results from J0034$-$0534 to 3GHz, the maximum rms introduced by the SW is about 65~ns, while the reported rms of the timing residuals of PSR~J1909$-$3744 is 200~ns \citep{rhc16}. \\ 
   The American PTA (the ``North American Nanohertz Observatory for Gravitational Waves'', \citealt{abb18}) eliminates ToAs when the predicted delay from a spherical model with $n_0=5$ (see Equation~\ref{eq:sphericalmodel}) exceeds 160 nanoseconds.\\ 
   The European PTA \citep{dcl16} does not exclude any data points and applies the T2 correction described in Section~\ref{sec:modeldescription} without the variable amplitude. \\ 
   
  A more systematic investigation on pulsars at different ecliptic latitudes is necessary before drawing final conclusions. However, the results shown in Figure~\ref{fig:rms} and the possibility that the SW induces false GWB detections \citep{thk16}, it is advisable to exclude ToAs closer than, at least, 20~degrees to the Sun. \\
 While the Y07 model appears to need further improvements to be successfully used to correct pulsar-timing residuals, it can be more usefully applied in simulations, as it represents the high-frequency temporal features of the SW much better than the T2 model. Possible evolutions of the Y07 model can be made, e.g., by searching for new scaling laws to describe the electron densities in the two SW phases, by introducing a time dependency of the slow-wind latitudinal extent (as it is expected to become thinner closer to the Solar minimum, and thicker -- but more mixed with the fast wind -- at the Solar maximum), or by introducing two different speeds for the fast and the slow phases.\\
 The spherical approximation, although it is more efficient, still leaves time-delay residuals up to 300~ns at 1.4 GHz within 20 degrees from the Sun (with the yearly-variable amplitude applied). In addition to this, we need to stress that both of the models were improved by a yearly-variable amplitude, that we were able to fit thanks to the exceptional cadence and precision of the GLOW data on PSR~J0034$-$0534.\\
 In case of high-precision, long-term pulsar-timing experiments with high-frequency datasets ($\sim$1400 MHz), the most efficient way to correct for SW and IISM effects at present, seems to be to perform simultaneous observations with low-frequency telescopes such as LOFAR, or, alternatively, to perform observations with wide-band receivers. However, caution should be taken with respect to the predicted phenomenon of frequency-dependent DM \citep{css16}, that might jeopardize the realization of this correction scheme. While an actual detection of this effect has been made in the case of a long-period pulsar \citep{dvt19}, it has never been detected for millisecond pulsars.

 \section{The Way Forward for Solar-wind Models in Pulsar Timing}
 
Our analysis has demonstrated that the currently available SW models do not adequately predict the electron-density variations that the SW induces in pulsar-timing data. This was expected, given the complexity of the heliosphere and the simplicity of the models.\\
Given the unprecedented sensitivity of our low-frequency data, we expect that with new and future telescopes it will be possible to turn our present work around, and develop more advanced SW models for pulsar timing purposes. Any such further development should properly deal with at least three components.\\

The first is the time-dependency of the spatial distribution of the SW electron density. While in our analysis this has been reached by performing a fit for a variable amplitude, such an approach might not be viable at higher observing frequencies, because of a non-sufficient sensitivity, see \citealt{mca19}.\\

The second component is the dependency of the SW electron density and velocity on the heliographic latitude and longitude (the latter due to non-dipolar components of the magnetic field). This is particularly important, e.g., around Solar minimum when the SW effectively becomes an oblate spheroid with significantly lower density over the magnetic poles, or in presence of variations that persist for several 27-day solar rotation periods, so that they are quasi-periodic \citep{par58}. The density variations are anti-correlated with the velocity (high velocity implies low density) and they do not persist as clearly as the velocity variations because of the interactions of fast and slow streams which build up with distance from the Sun (the already mentioned CIRs). These fast and slow streams cannot penetrate each other so the fast streams are deflected away from the direction of rotation and the slow streams are deflected the other way. It is impossible to model this with a pure kinematic code since the CIRs are highly turbulent; but it might be possible with a full 3D magneto-hydrodynamic simulation. The Y07 model uses observations of the Solar magnetic field to attempt a prediction of the SW electron density, accounting for this non-stationary component. As we have shown, its predictions are still underestimations of the true density variations, but the order of magnitude appears correct. Further developments of Y07 model may improve on this situation, for example by varying the velocities of the SW components, integrating along the line of sight in smaller segments, fitting a continuously (but slowly) varying SW amplitude, etc. At the very least this should be able to provide a model that can accurately predict the statistics of the quasi-stationary components, even if it cannot predict their positions with sufficient accuracy to allow correcting high-precision pulsar-timing data.\\

The third and final component of a SW model would address coronal mass ejections (CMEs) which occur primarily at mid-latitudes and propagate outwards as independent plasmoids. CMEs often move faster than the SW into which they are injected and consequently decelerate significantly by the time they reach terrestrial orbit. Prediction of the effects of CIRs and CMEs at the Earth is part of ``space weather'' research programmes, but these are not yet reliable enough to predict the electron density at Earth.\\

Low-frequency radio observations are highly sensitive to propagation effects and provide a valuable basis on which to further develop and expand the existing SW models. However, as we have shown, this is a non-trivial experiment not only because of the SW complexity, but also because of the complexity of the data themselves, which are bound to also show variations due to interstellar structures. In order to separate these two effects we have employed an ad-hoc method that does not rely on any prior assumptions but a more ideal approach would construct a Wiener filter to optimally disentangle the two processes. The disadvantage of a Wiener-filter approach in disentangling the SW from the IISM, is that it requires \textit{apriori} knowledge of the power spectra of both processes. For the IISM component statistical models exist (the see, e.g., \citealt{kcs13}), but for the SW this was not the case. Based on the research presented in
this article, however, it appears that the Y07 model may be suitable as the basis for this, given that its predictions for the covariance of the SW impact on pulsar-timing residuals is of the correct order of magnitude. On the basis of this, therefore, a Wiener filter can be constructed to optimally disentangle SW and IISM contributions in pulsar-timing data, but the development and application of such method is beyond the scope of this article.

 \section{Summary and conclusions}\label{sec:conclusions}
 
We have used high-cadence, low-frequency observations of PSR~J0034$-$0534 to study the SW effects on the DM time series in detail, and to test the performance of the mitigation routines that are currently available in pulsar timing. This has shown that for the data set considered in our work:

\begin{itemize}
\item The default, spherical model in pulsar timing (as described by \citealt{ehm06} but improved with a yearly-variable amplitude) is insufficient for current high-precision timing requirements in the analysed pulsar, as it leaves ToA corruptions at levels beyond 100 ns at 1400~MHz in observations up to 40 degrees away from the Sun. The problem of having good SW models will be of a particular importance for timing programs carried out with future, more sensitive facilities such as the Square Kilometer Array \citep{bbg15} or the Five hundred meter Aperture Spherical Telescope \citep{pns01};
\item The two-phase model introduced by \citet{yhc07}, which was also improved with a yearly-variable amplitude as introduced in \citet{ych12}, performs worse at almost all angular distances between the Sun and the pulsar than the spherical model. That is, the post-correction ToAs contain more noise than those that are corrected by the spherical Solar-wind model;
\item The two-phase model of \citealt{yhc07} does provide a reasonable approximation for the impact of high-frequency temporal SW fluctuations on ToAs, so it can be used for simulation purposes. However, it must be noted that it provides a slightly underestimated prediction of the SW effect.
\end{itemize}

Based on the above, it is clear that neither of the presently-available SW models are sufficient to correct the PTA pulsar that we analysed at a 100-ns level when the angular separation between the Sun and the pulsar is less than 40~degrees. If this is the case for a number of other PTA pulsars, high-precision pulsar-timing experiments should consider a number of alternative options. Moving the timing to higher frequencies would reduce the SW impact, but may negatively affect timing precision due to the typically steep spectral indices of pulsars \citep{blv13}. Simultaneous low-frequency data (like those used in the present article) can be used to provide independent estimates of the SW density and thereby correct the higher-frequency timing data. This approach has the disadvantage that the observations will need to be exactly simultaneous given the often short-lived structures in the SW. So-called frequency-dependent DMs \citep{css16, dvt19} are unlikely to pose a problem in this context due to the relative proximity of the SW to Earth. Ultra-broad-band receiver systems will also be able to simultaneously provide highly accurate estimates of the SW density and highly precise ToAs \citep{pen19,ldc14}. Finally, there are multiple possible ways in which to further develop the current SW models, to improve their assumptions and test their predictions -- particularly on the basis of highly sensitive low-frequency data as used in this article. Such further development, beyond the scope of this article, would require a larger set of pulsars across various ecliptic latitudes and a multi-wavelength, inter-disciplinary approach. Such an analysis will provide useful insights into space weather and Solar physics and will lead to new and improved ways to mitigate SW effects in pulsar-timing experiments. Space weather studies will also benefit from the increase of the (currently) scarce statistics of magnetic-field measurements in the SW, as this will improve the modelling capability of the magnetic field in the heliosphere.
 
\section*{Acknowledgements}
CT is grateful to William Coles for the large amount of inputs that he gave to this work, and for his mentorship and all the knowledge that he shared with the authors. This article is based on data obtained with the German stations of the International LOFAR Telescope (ILT), constructed by ASTRON \citep{vwg13}. The observations used in this work were made during station-owners time as well as during ILT time allocated under the long term pulsar monitoring campaign project (PI: M. Serylak) with codes: LC0\_014, LC1\_048, LC2\_011, LC3\_029, LC4\_025 and LT5\_001. In this work we made use of data from the Effelsberg (DE601) LOFAR station funded by the Max-Planck-Gesellschaft; the Unterweilenbach (DE602) LOFAR station funded by the Max-Planck-Institut f\"ur Astrophysik, Garching, and the Max-Planck-Gesellschaft; the Tautenburg (DE603) LOFAR station funded by the State of Thuringia, supported by the European Union (EFRE) and the Federal Ministry of Education and Research (BMBF) Verbundforschung project D-LOFAR I (grant 05A08ST1); the J\"ulich (DE605) LOFAR station supported by the BMBF Verbundforschung project D-LOFAR I (grant 05A08LJ1). The observations of the German LOFAR stations were carried out in the stand-alone GLOW mode (German LOng-Wavelength array), which is technically operated and supported by the Max-Planck-Institut f\"ur Radioastronomie, the Forschungszentrum J\"ulich and Bielefeld University. We acknowledge support and operation of the GLOW network, computing and storage facilities by the FZ-J\"ulich, the MPIfR and Bielefeld University and financial support from BMBF D-LOFAR III (grant 05A14PBA), and by the states of North Rhine-Westphalia and Hamburg. SO acknowledges support from the Alexander von Humboldt Foundation and the Australian Research Council Laureate Fellowship grant FL150100148. The authors thank James McKee for confirming the DM structure function of the GLOW data.




\bibliographystyle{mnras}


%
%


\bsp	
\label{lastpage}

\end{document}